\documentclass[preprint,showpacs,pre,12pt]{revtex4-1}
\usepackage{amsfonts}
\usepackage{amssymb}
\usepackage{amsmath}
\usepackage{graphicx}
\usepackage{pgfplots}

\begin{document}
\title{Level compressibility  of certain  random unitary matrices}
\author{Eugene Bogomolny}
\affiliation{Universit\'e Paris-Saclay, CNRS, LPTMS, 91405 Orsay, France}
\date{\today}
\begin{abstract}
The value of spectral form factor at the origin, called level compressibility,  is an important characteristic 
of random spectra. The paper is devoted to analytical calculations of this quantity for different  random unitary matrices describing models  with intermediate spectral statistics. The computations are based on the approach developed  by G. Tanner in [J. Phys. A: Math. Gen. \textbf{34},  8485 (2001)] for chaotic systems. The main ingredient of the method is the determination of eigenvalues of a transition matrix whose matrix elements equal squared moduli   of  matrix elements of the initial unitary matrix. The principal result of the paper is the proof that the level compressibility of random unitary matrices derived from the exact quantisation of barrier billiards and consequently of barrier billiards themselves is equal to $1/2$ irrespectively of the height and the position of the barrier.    
\end{abstract}

\maketitle

%================================================
\section{Introduction}

The leading idea behind statistical descriptions of   complex deterministic quantum problems is  that quantum characteristics  (e.g., eigenenergies) of a large variety of such  problems  are so erratic and irregular that their precise values are irrelevant (like the position of a molecule in the air) and only their statistical properties  are of importance. As matrices are inherent in quantum mechanics, random matrices occupy a predominant place in the  application of statistics to quantum problems \cite{wigner}. In a typical  setting one tries to find a random matrix ensemble whose eigenvalues have the same statistical distributions as (high-excited) eigenenergies of a given deterministic quantum problem. Till now this query has been  figured out  only for two limiting classes of  quantum problems:   (i)  models whose classical limit is integrable \cite{berry_tabor} and (ii) models whose classical limit is chaotic \cite{bgs}. For  generic integrable models quantum eigenenergies  are distributed as eigenvalues  of diagonal matrices with independent identically distributed (i.i.d.) elements which means that their correlation functions after unfolding coincide with the ones of the  Poisson distribution \cite{berry_tabor}. For generic chaotic systems   it was conjectured in \cite{bgs} that their eigenenergies are distributed as eigenvalues of standard random matrix  ensembles (GOE, GUE, GSE) depending only on system symmetries whose correlation functions are known explicitly \cite{mehta}.  The difference between these two cases is clearly seen from the limiting behaviour of their nearest-neighbour distribution $P_0(s)$ which is the probability density that two eigenvalues are separated by a distance $s$ and there are  no other eigenvalues in-between. For the Poisson statistics there is no level repulsion which means that $\lim_{s\to 0}P_0(s)\neq 0$  and for large argument $P_0(s)$ decreases exponentially with $s$. For standard random matrix ensembles levels repel each other, $\lim_{s\to 0}P_0(s)=0$,  and $P_0(s)\sim \exp(-as^2)$ when $s\to\infty$.   

These two big conjectures form a cornerstone of quantum chaos and have been successfully applied to various problems from nuclear physics to number theory. Nevertheless, they do not cover all possible types of models. 
Especially intriguing is the class of pseudo-integrable billiards (see, e.g., \cite{richens_berry}) which are 2-dimensional polygonal billiards whose angles $\theta_j$ are rational multiplies of $\pi$
\begin{equation*}
\theta_j=\frac{m_j}{n_j}\pi 
\end{equation*}
with co-prime integers $m_j$ and $n_j$. A peculiarity of such billiards is seen  in the fact that their classical trajectories belong to a 2-dimensional surface of genus $g$ related with angles as follows \cite{katok}
\begin{equation*}
g=1+\frac{N_n}{2}\sum_j\frac{m_j-1}{n_j}
\end{equation*}  
where $N_n$ is the least common multiple of all denominators $n_j$. Consequently, any such model with at least one numerator $m_j>1$ is neither  integrable (which would imply that  trajectories belong to a 2-dimensional torus with $g=1$) nor fully chaotic (in which case trajectories should cover 3-dimensional surface of constant energy) and   the aforementioned  conjectures cannot be applied to such systems. Numerical calculations show that for many pseudo-integrable billiards  spectral statistical properties of corresponding quantum problems differ from both the Poisson statistics  and the random matrix statistics mentioned above (see, e.g., \cite{billiard_1, billiard_2} and references therein). In particular, for these models (i) $\lim_{s\to 0} P_0(s)=0$ as for standard random matrix ensembles but (ii) $P_0(s)\sim \exp(-bs) $ for large $s$ as for the Poisson statistics.  Such hybrid statistics, labeled  intermediate statistics, had been first observed in the Anderson model at the metal-insulate transition \cite{altshuler, shklovskii} and they constitute a special, interesting but  poorly  investigated class  of spectral statistics. 

\begin{figure}
\begin{minipage}{.49\linewidth}
\vspace{10pt}
\begin{center}
\includegraphics[width=.99\linewidth]{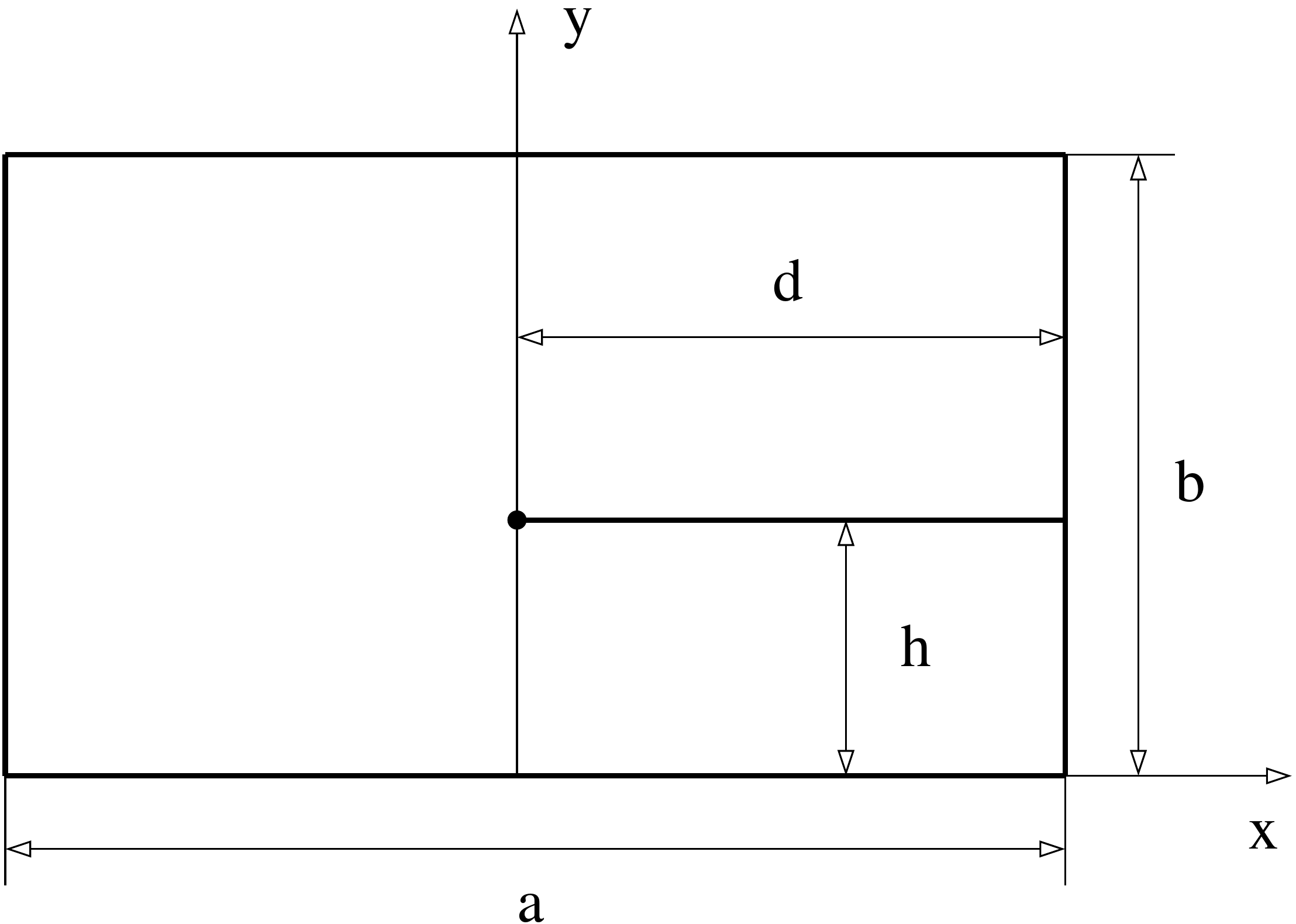}\\
(a) 
\end{center}
\end{minipage}
\begin{minipage}{.49\linewidth} 
\begin{center}
\includegraphics[width=.93\linewidth]{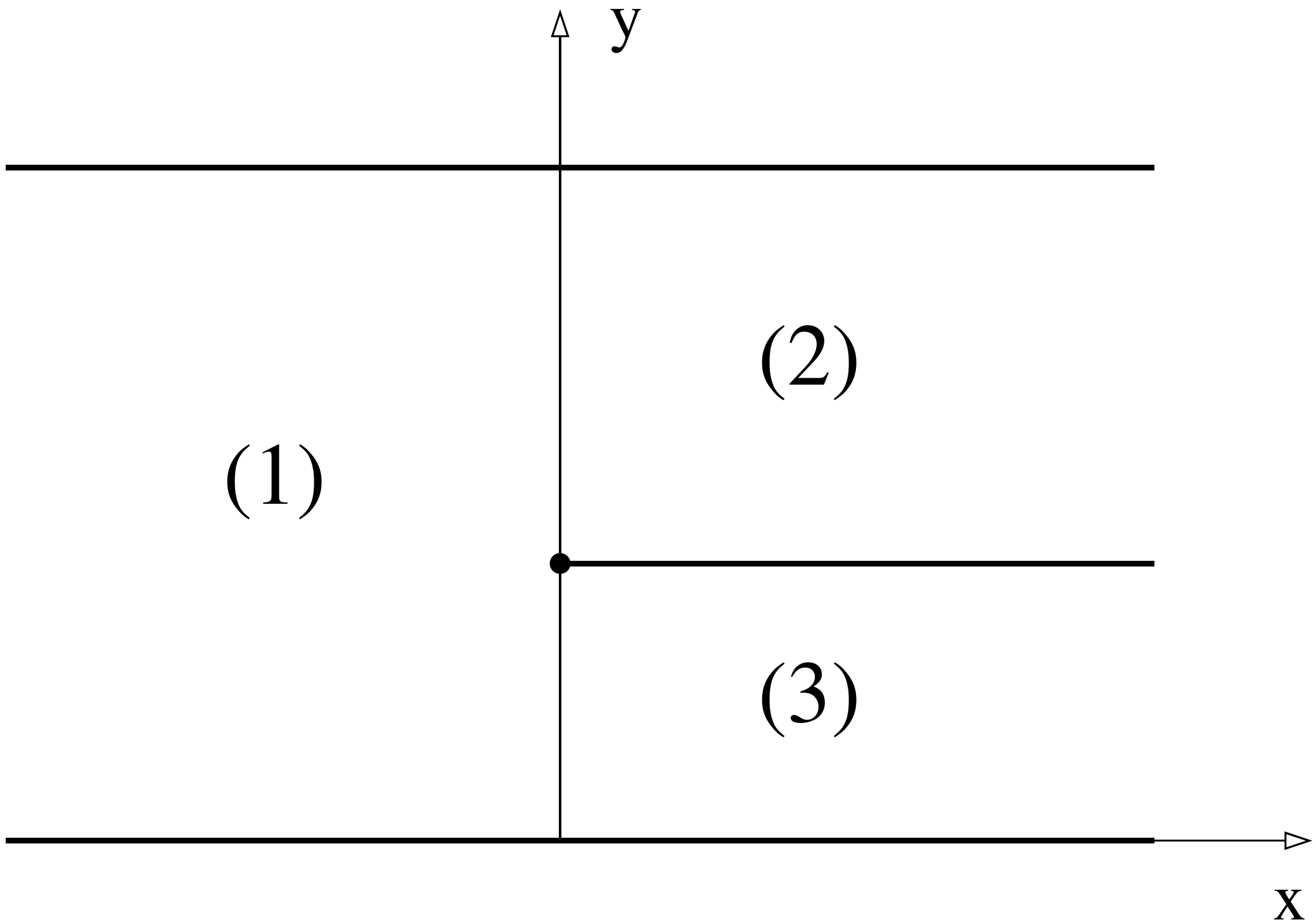}\\
(b)
\end{center}
\end{minipage}
\caption{(a) Barrier billiard. (b) An infinite  slab with a half-plane inside. Numbers indicate 3 possible channels.} 
\label{barrier_fig}
\end{figure}

Probably the  simplest  example of pseudointegrable systems  is the so-called barrier billiard   which is a rectangular billiard with a barrier inside sketched in figure~\ref{barrier_fig}(a).   The quantum problem for this model consists in solving the Helmholtz equation  
\begin{equation*}
(\Delta +E_{\alpha})\Psi_{\alpha}(x,y)=0
\end{equation*} 
imposing that eigenfunction $\Psi_{\alpha}(x,y)$ obeys (e.g.) the Dirichlet  conditions on the boundary of the rectangle as well as on the barrier
\begin{equation*}
\Psi_{\alpha}(x,y)|_{\mathrm{boundary}}=0,\qquad \Psi_{\alpha}(x,y)|_{\mathrm{barrier}}=0.
\end{equation*} 

Calculating the exact $S$-matrix for the scattering inside the infinite slab with a barrier depicted in figure~\ref{barrier_fig}(b),  it has been demonstrated in \cite{billiard_1, billiard_2}  that spectral statistics of this model is the same as the statistics of eigenvalues of the following $N\times N$ random unitary matrix 
\begin{equation}
B_{m,n}=  e^{i\Phi_m} \frac{L_m L_n}{x_m+x_n}, \qquad m,n=1,\ldots, N
\label{B_matrix}
\end{equation}
where $\Phi_m$ are i.i.d. random variables uniformly distributed on interval $[0,2\pi)$ and 
\begin{equation}
L_m=\sqrt{2x_m\prod_{k\neq m}\frac{x_m+x_k}{x_m-x_k}}
\label{L_m}
\end{equation}
where coordinates $x_m$  depend on the position of the barrier. 

Define the following quantities (momenta) of propagating modes in each of 3 channels indicated in figure~\ref{barrier_fig}(b)
\begin{equation*}
p_m^{(1)}=\sqrt{k^2-\frac{\pi^2 m^2}{b^2}}, \qquad  p_m^{(2)}=\sqrt{k^2-\frac{\pi^2m^2}{(b-h)^2}},\qquad p_m^{(3)}=\sqrt{k^2-\frac{\pi^2 m^2}{h^2}}
\end{equation*}
If  the ratio $h/b$ is an irrational number, coordinates $x_m$ have the following form 
\begin{equation}
\vec{x}=b\Big( \underbrace{ p_1^{(1)},\ldots,  p_{N_1}^{(1)}}_{\text{$N_1$}}, \underbrace{-p_1^{(2)},\ldots, - p_{N_2}^{(2)}}_{\text{$N_2$}},\underbrace{-p_1^{(3)},\ldots,  -p_{N_3}^{(3)}}_{\text{$N_3$}}\Big ).
\label{x_irrational}
\end{equation}
Here $N_j$ with $j=1,2,3$ are the numbers of propagating modes in each channel
\begin{equation}
N_1=\left [ \frac{kb}{\pi}\right ],\qquad N_2=\left [ \frac{k(b-h)}{\pi}\right ],\qquad N_3=\left [ \frac{kh}{\pi}\right ]
\label{N_irrational}
\end{equation}
where $[x]$ is the largest integer $\leq x$ and the total dimension of the $B$-matrix is $N=N_1+N_2+N_3$.

When the ratio $h/b$ is a rational number, $ h/b=p/q$ with co-prime integers $p$ and $q$ ($p<q$),  there are exist exact   plane wave solutions of barrier billiard equal zero at the whole line passing through the barrier. It is natural to disregard them and take into account only non-trivial eigenvalues. In such case coordinates $x_m$ have to be chosen as indicated below 
\begin{equation}
\vec{x}=b\Big( \underbrace{ p_1^{(1)},\ldots,p_k^{(1)},\ldots,  p_{N_1}^{(1)}}_{\text{$k\neq 0 \mod q$}}, \underbrace{-p_1^{(2)},\ldots, - p_{N_2}^{(2)}}_{\text{$N_2$}},\underbrace{-p_1^{(3)},\ldots,-p_k^{(3)},\ldots,   -p_{N_3}^{(3)}}_{\text{$k\neq 0 \mod p $}}\Big )\, .
\label{x_m_q}
\end{equation}
The dimension of this  vector is 
\begin{equation}
N^{(\mathrm{r})}=N_1+N_2+N_3-2N_0,\qquad  N_0=\left [\frac{kb}{\pi q} \right]  .
 \label{N_m_q}
 \end{equation}
The matrix $B$ can be generalised for arbitrary vector $x_m$ provided the following interlacing conditions are fulfilled 
\begin{equation*}
|x_1|>|x_2|>\ldots >|x_N|, \qquad x_m=(-1)^{m+1}|x_m| .
\end{equation*}
Exact correlation functions for the $B$-matrix are unknown at present. In \cite{billiard_1, billiard_2} it was argued that an approximate  Wigner-type surmise for this matrix corresponds to the so-called semi-Poisson distribution \cite{gerland}. In particular,  it implies that the probability density $P_n(s)$ that two levels are separated by a distance $s$ and there are exactly $n$ levels in-between (after the standard unfolding)  is given by the following expression
\begin{equation*}
P_n(s)=\frac{2^{2n+2}}{(2n + 1)!}s^{2n+1}e^{-2s},\qquad n\geq 0\, .
\end{equation*}
Numerical calculations presented in \cite{billiard_1, billiard_2} agree with  these simple formulas. 

This paper is devoted to the calculation of another important characteristic of spectral statistics, namely the level (or spectral)  compressibility.  This quantity is determined by the limiting behaviour of the variance of the number of levels inside a given interval. More precisely, let $N(L)$ be the number of eigenvalues in an interval $L$ unfolded to the unit mean density which means that  the mean number of levels in interval $L$ equals $L$,
$\langle N(L)\rangle = L $. By  definition the number variance is  
$\Sigma^{(2)}(L)\equiv \langle (N(L)-L)^2\rangle$. If for large $L$ 
\begin{equation}
\Sigma^{(2)}(L)\underset{L\to\infty}{\longrightarrow} \chi L
\label{chi}
\end{equation}
constant $\chi$ is called the level  compressibility. The importance of this quantity follows from the fact  that for integrable systems with the Poisson statistics $\chi=1$ but for standard random matrix statistics typical for chaotic models $\chi=0$. For all examples of intermediate statistics it was observed that $0<\chi<1$. 

The conventional way of determination the level compressibility for dynamical systems is the summation over all periodic orbits in the diagonal approximation  initiated in \cite{rigidity}.  For the symmetric barrier billiard with $h/b=1/2$ and $d/a=1/2$  it has been demonstrated in \cite{wiersig} that $\chi=1/2$. The same value had been obtained in \cite{olivier_thesis}  for the case $h/b=1/2$ and arbitrary barrier height  $d/a$.  Finally  for $h/b=p/q$ with co-prime integers $p$, $q$ and irrational values of $d/a$ it has been proved in \cite{giraud}  that 
\begin{equation*}
\chi=\frac{1}{2}+\frac{1}{q}
\end{equation*}
but in the calculations  exact eigenvalues whose eigenfunctions are zero on the whole line $y=h$ (see figure~\ref{barrier_fig})(a) have not been excluded. When these trivial eigenvalues are removed the answer is $\chi=1/2$ \cite{private_communication}. Therefore direct (and quite tedious) calculations suggest  that for (almost) all positions and heights of the barrier the level compressibility is the same as for the semi-Poisson statistics \cite{gerland}
\begin{equation}
 \chi=\frac{1}{2} 
 \label{chi_billiard}
\end{equation}
but the reason of this universality remains obscure.  

The purpose of the paper is to find analytically the spectral compressibility for  barrier billiards and for a few other models directly from the corresponding random unitary  matrices.  To achieve the goal it is   convenient to slightly generalise the  method developed  in \cite{tanner} for random unitary matrices appeared in the quantisation of quantum graphs \cite{gnutzmann}. The method is briefly explained in Section~\ref{generalities}. In Section~\ref{old_map} this method is applied to random unitary matrices derived in \cite{marklof}  by quantisation of a simple  interval-exchange map.  In this case the transition matrix is a circulant matrix whose eigenvalues are known explicitly.  The results coincide  with exact level compressibility for these models obtained in \cite{map}-\cite{integrable_ensembles}.  This example gives credit to the method and permits to explain its main features without unnecessary complications.  In particular it clarifies the situation (not covered by Refs.~\cite{map}-\cite{integrable_ensembles}) when a  parameter entered the matrix takes an irrational value. Numerically it has been observed \cite{general_case} that in such case the spectral statistics are well described  by  the ones of chaotic systems (GOE or GUE) though the Lyapunov exponent of the underlying classical map is always zero.      

The main part of the paper is devoted to the derivation of the level compressibility for random matrices associated with barrier billiards \eqref{B_matrix}. The calculation are more complicated as no eigenvalues (except  one) are known analytically.  The simplest case of  the symmetric barrier billiard with $h/b=1/2$ is investigated in Section~\ref{symmetric_BB}.  To get tractable expressions a kind of paraxial approximation is developed which permits to control the largest terms. By using such approximation the transition matrix is transformed into a Toeplitz matrix with a quickly decreasing symbol which allows  to find its eigenvalues for large matrix dimension.  The result of this Section is that the level compressibility equals $1/2$ in accordance with periodic orbit calculations in \cite{wiersig} and \cite{olivier_thesis}. 

In   Section~\ref{irrational_BB} random unitary matrices corresponding to barrier billiards with irrational ratio $h/b$ (with  coordinates given by \eqref{x_irrational}) are considered.  In this case the transition matrix in paraxial approximation contains quickly oscillating  terms and, consequently, has forbidden zones in the spectrum. In spite of that  one can argue that largest moduli eigenvalues are insensitive to fast oscillations and are determined   solely by a matrix averaged over such oscillations.  The final matrix is also a Toeplitz  type which allows of analytical calculations proving that the level compressibility is again equal $1/2$.    

Section~\ref{rational_BB} is  addressed to the calculation of level compressibility in the most complicated case of barrier billiards with rational ratio $h/b=p/q\neq 1/2$, the only  one for which the direct calculation of the level compressibility by the summation over periodic orbits was not yet done. The computation is  cumbersome but in the end one comes to the conclusion that the level compressibility remains equal $1/2$. In other words the level compressibility of  barrier billiards  is universal (i.e., independent of the barrier position and its height)  and coincides with the semi-Poisson prediction \cite{gerland}. 

Section~\ref{conclusion} gives a brief summary of the results. A few technical points are discussed in Appendices~\ref{direct_calculation}-\ref{divers}.        

%------------------------------------- 
\section{Generalities}\label{generalities}

It is well known that the level compressibility \eqref{chi} is related with  the two-point correlation form factor $K(\tau)$  as follows
\begin{equation}
\chi=\lim_{\tau\to 0} K(\tau)\, .
\label{limit_K}
\end{equation} 
For $N\times N$ random unitary matrices $U$ the form factor  can, conveniently, be written in the following  concise form (see, e.g.,\cite{tanner}) 
 \begin{equation}
K(\tau)= \Big \langle \frac{1}{N} \left | \mathrm{Tr}\, U ^{n}\right |^2 \Big \rangle , \qquad \tau=\frac{n}{N}
\label{K_trace}
\end{equation}
where the average is taken either on different realisations of random parameters, or over of a small window of $\tau$, or the both. 

Unitary matrices considered in the paper all have the product form 
\begin{equation}
U_{j,k}=e^{i\Phi_j} w_{j,k}
\label{main_matrix}
\end{equation}
where $\Phi_j$ are i.i.d.  random phases uniformly distributed between $0$ and $2\pi$ and matrix $w_{j,k}$ is a fixed unitary matrix. 

In \cite{tanner} (see \cite{gnutzmann} for more detailed discussion) it was shown that for such  unitary matrices the averaging over random phases   leads in the diagonal approximation to the following formula
\begin{equation}
K^{(\mathrm{diag})}\Big (\frac{n}{N}\Big )=g\frac{n}{N}\mathrm{Tr}\,  \left (T^{n} \right )
\label{K_tau}
\end{equation}
where matrix elements of matrix $T$, called below the transition matrix, are squared moduli  of matrix elements of matrix $U$ 
\begin{equation}
T_{j,k}=|U_{j,k}|^2=|w_{j,k}|^2\, .
\label{tanner_matrix}
\end{equation}
For systems without time-reversal invariance $g=1$  and for models with time-reversal invariance  $g=2$ (here only cases of $g=1$ are considered). Due to the unitarily of matrix $U$ the $T$-matrix is double stochastic matrix, $\sum_j T_{j,k}=1$ and $\sum_k T_{j,k}=1$,  thus having  the meaning of classical transition matrix.  

Let $\Lambda_{\beta}$ be eigenvalues of the $T$-matrix
\begin{equation*}
T_{j,k} u_k(\beta)=\Lambda_{\beta} u_j(\beta),\qquad  \beta=0,\ldots,N-1\, .
\end{equation*}
From \eqref{K_tau} it follows that 
\begin{equation}
K^{(\mathrm{diag})}\Big (\frac{n}{N}\Big )=\frac{n}{N}\sum_{\beta=0}^{N-1}  \Lambda_{\beta}^n \, .
\label{K_lambda} 
\end{equation}
The unitarity imposes that one eigenvalues $\Lambda_0=1$. Perron-Frobenius theorem states that all other eigenvalues $|\Lambda_{\beta}|\leq 1$. Let the set of eigenvalues be ordered
$1=\Lambda_0\geq \Lambda_1\geq \ldots \geq \Lambda_{N-1} $. 
 
One has 
\begin{equation}
K^{(\mathrm{diag})}(\tau)=\tau\left (\mathrm{Tr}\,  \left (T^{n} \right )\right ) =
\tau +\tau \sum_{\beta=1}^{N-1}\Lambda_{\beta}^{N \tau},\qquad 
\sum_{\beta=1}^{N-1}\Lambda_{\beta}^{N \tau}
 \leq  (N-1)|\Lambda_1|^{N\tau}\, . 
\label{tanner_inequality} 
\end{equation}
It has been noted  in \cite{tanner} that if 
\begin{equation}
\lim_{N\to\infty} |\Lambda_1|^{N}=0
\label{tanner_condition}
\end{equation}
then the second term in \eqref{tanner_inequality} goes to zero for all finite $\tau$ and  $K^{(\mathrm{diag})}(\tau)=\tau$ for small $\tau$. Consequently  it is reasonable  to conjecture (as it has been done in \cite{tanner})  that the whole spectral statistics of such matrices will  be well described by standard random matrix formulas. 

For matrices discussed below  criterium  \eqref{tanner_condition} is not fulfilled. Instead, in all considered cases largest moduli eigenvalues of transition matrices  have the form $1-\mathcal{O}(1/N)$. To calculate  the level compressibility from \eqref{K_lambda}  the summation over all such  eigenvalues  is performed analytically and then the limit $\tau\to 0$ is taken. 

 It is well known that the form factor is not a self-average quantity. It has strong fluctuations and necessarily requires a smoothing. There exist two different sources of fluctuations. The first is related with random  phases in  matrices \eqref{main_matrix}.  Eq.~\eqref{K_tau}   corresponds to the averaging over these random phases in the diagonal approximation.  The second has its roots in non-smoothness of $K^{(\mathrm{diag})}(n/N)$ for different $n$ and could be removed by a smoothing over a small interval of $\tau$. (A trivial example is $(-1)^n$.)
 
%================================================
\section{Interval-exchange matrices}\label{old_map}

This Section is devoted to the calculation of the level compressibility for special unitary matrices derived in \cite{marklof} by quantisation of a simple 2-dimensional parabolic map. Slightly generalising their result \cite{map} one can write these  matrices in  the following form
\begin{equation}
M_{n,m}=e^{i\Phi_n} \frac{1-e^{2\pi i \alpha N }}{N\big (1-e^{2\pi i(n-m+\alpha N )/N}\big )} ,\qquad n,m=1,\ldots, N.  
\label{map}
\end{equation}
Here $\alpha$ is a real parameter and  $\Phi_n$ are i.i.d. random variables uniformly distributed between $0$ and $2\pi$ (the case with 'time-reversal  symmetry' when $\Phi_{N-n+1}=\Phi_n$  requires only multiplication the formulas below  by factor $g=2$ as indicated  in \eqref{K_tau}).    

When $\alpha$ is a rational number $\alpha=p/q$ with co-prime integers $p$ and $q$ the original classical map is an interval-exchange map and, as it was shown in detail in  \cite{map}-\cite{integrable_ensembles},   spectral statistics of matrices $\eqref{map}$ in the limit of $N\to\infty$ are unusual and peculiar.  It appears that the limiting results depend on the residue of $p N \mod q$ (when $pN\equiv 0 \mod q$ matrix \eqref{map} have explicit eigenvalues not interesting for our purposes).  For example,  if $\alpha=1/5$ there are 2 possibilities:   $N\equiv \pm 1$ mod $5$  and $N\equiv \pm 2\mod 5 $. In the first case  the nearest-neighbour distribution is 
\begin{equation*}
P_0(s)=\frac{5^5}{4!}s^4 e^{-5s}
\end{equation*} 
but for the second one  the exact result is different
\begin{equation*}
P_0(s)=(a_2 s^2 + a_3 s^3 + a_4 s^4 + a_5 s^5 + a_6 s^6)e^{-5s} 
\end{equation*}
where $a_2 = 625/2-275\sqrt{5}/2$, $a_3 = 3125/2-1375\sqrt{5}/2$, $a_4 = 71875/48+ 33125\sqrt{5}/48$, 
$a_5=-15625/3 + 9375\sqrt{5}/4$, $a_6 = 1015625/288-453125\sqrt{5}/288$.

Though the spectral correlation functions for $\alpha=p/q$ are different for different residues $pN \mod q$, the calculations show that the spectral compressibility for all residues  remains  the same \cite{map, general_case}
\begin{equation}
\chi=\frac{1}{q}\, . 
\label{chi_rational}
\end{equation} 
Matrices \eqref{map} with  irrational $\alpha$ were investigated  numerically in \cite{general_case}  and it was observed that their spectral statistics are well described by standard random matrix ensembles (GOE or GUE). In particular, it implies that in such case 
\begin{equation}
\chi=0\, . 
\label{chi_irrational}
\end{equation} 
Below it is shown that values \eqref{chi_rational} and \eqref{chi_irrational} can easily  be recovered by the discussed method. 

The transition matrix \eqref{tanner_matrix} for the discussed case has the form   
\begin{equation}
T_{n,m}\equiv |M_{n,m}|^2= \frac{\sin^2 (\pi \alpha N )}{N^2 \sin^2 \big (\pi (n-m+\alpha N )/ N\big )}
\label{T_interval_exchange}
\end{equation}
This  is a circulant matrix and its eigenvalues are simply  the Fourier transforms of its matrix elements 
\begin{equation}
\Lambda_{\beta}=\frac{\sin^2 (\pi \alpha N )}{N^2} \sum_{s=0}^{N-1}  \frac{ e^{-2\pi i \beta s /N }}{\sin^2 \big (\pi (s+\alpha N )/ N\big )}
\label{eigenvalues_T}
\end{equation}
Differentiating the  both sides of the identity with integer $s$ on $z$  
\begin{equation*}
\sum_{m=0}^{N-1} e^{2\pi i m(s+z)/N}=\frac{1-e^{2\pi i z}}{1-e^{2\pi i (s+z)/N}},
\end{equation*}
after straightforward  transformations one proves  that
\begin{equation*}
 \frac{ \sin^2(\pi z) }{ \sin^2\big (\pi (s+z)/N\big )}=
-2i \sin(\pi z) e^{-\pi i z} \sum_{m=0}^{N-1}\left ( m - N \frac{e^{\pi i z}}{2i \sin (\pi  z)}\right ) e^{2\pi i m(s+z)/N}
 \end{equation*}
 from which is follows  that eigenvalues \eqref{eigenvalues_T} are 
 \begin{equation}
 \Lambda_{\beta}=\left ( 1-\frac{\beta}{N}\Big (1-e^{-2\pi i \alpha N }\Big ) \right)  e^{2\pi i \beta \alpha },\qquad \beta=0,1\,\ldots,N-1 \, . 
 \label{Lambda_interval_exchange}
 \end{equation}
 Notice that $ \Lambda_0=1$, $\Lambda_{N-\beta}=\Lambda_{\beta}^{*}$ for $\beta=1,\ldots,N-1$. 
 
 Consider first the case of rational  $\alpha=p/q$.  Assume that $N\neq 0$ mod $q$ and calculate the form factor from \eqref{K_lambda} separately for $n\equiv r$ mod $q$ with  $r=0,1,\ldots,q-1$ (i.e., $n=qt+r$ with integer $t$) 
\begin{equation*}
K\left (\frac{q t+r}{N}\right ) =\frac{qt+r}{N} \left [1+2\mathrm{Re}\, \sum_{\beta=1}^{N/2} \left ( 1-\frac{\beta}{N}\Big (1-e^{-2\pi i \alpha N}\Big ) \right)^{qt+r}   e^{2\pi i \beta p r/q} \right ] .
\end{equation*}
Here we do not take into account that when $N=$ even the term with $\beta=N/2$ is real and has additional factor  $1/2$. 

Notice that the phase depends only on $r$. For large $N$ one can put the first factor in the exponent and sum the geometric progression from 1 to infinity. The answer is 
\begin{equation}
K_r(\tau)=\tau +2\tau \mathrm{Re}\left ( \frac{e^{-\xi  \tau +2\pi i p r/q}}{1-e^{-\xi  \tau +2\pi i p r/q}}\right ), \qquad \tau=\frac{qt+r}{N}
\label{K_r}
\end{equation} 
where $\xi=1-e^{-2\pi i p N/q }=1-e^{-2\pi i k/q}$ with $k=p N \mod q$. 

When $\tau\to 0$ all terms except the one with $r=0$ tends to zero as they do not have a pole at $\tau=0$. The remaining term equals
\begin{equation*}
\lim_{\tau\to 0} K_0(\tau)=2 \,\mathrm{Re} \left ( \frac{1}{\xi}\right ) =\frac{2\, \mathrm{Re}\,(1-e^{2\pi i z})}{2-2\cos(2\pi z)}=1\, . 
\end{equation*}
It means that after the averaging over random phases the limiting value of the form factor $K_r(\tau)$ strongly depends on small changes of $\tau=(qt+r)/N$. For  $r=0$ $K_0(\tau)$ tends to $1$ for small $\tau\ll1 $  but  all other terms with $r=1,\ldots,q-1$ $K_r(\tau)$ tend to $0$. The difference between these different values of $\tau$ is very small,  of the order of $1/N$. Therefore   after the averaging over any small (but finite) interval of $\tau$ one gets
\begin{equation*}
\lim_{\tau\to 0} \langle K(\tau)\rangle =\frac{1}{q}
\label{K_q}
\end{equation*}
which agrees with \eqref{chi_rational}  obtained in \cite{general_case} by a different method. 

\begin{figure}
\begin{minipage}{.49\linewidth}
\begin{center}
\includegraphics[width=.99\linewidth]{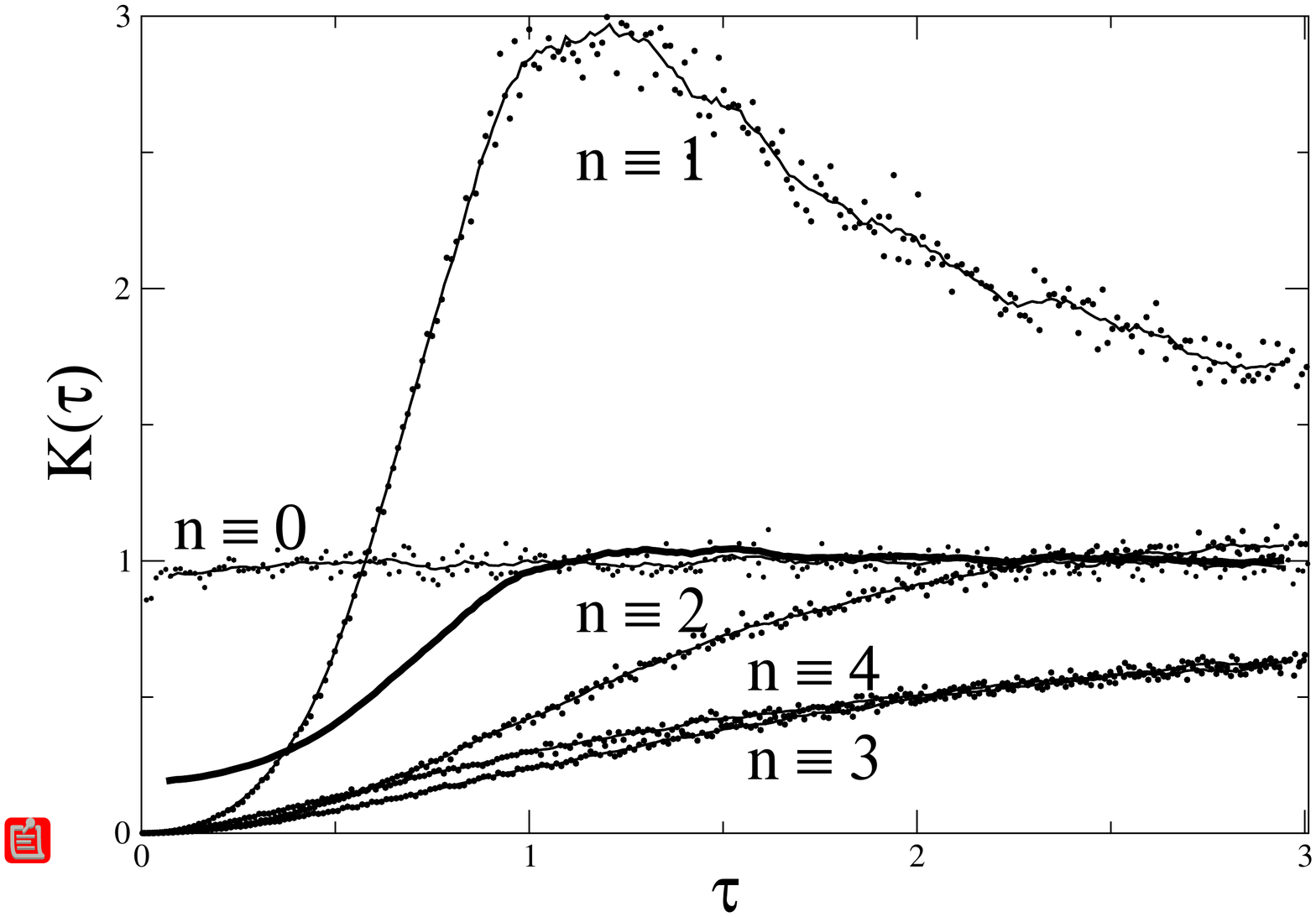}\\
(a)
\end{center}
\end{minipage}
\begin{minipage}{.49\linewidth}
\begin{center}
\includegraphics[width=.99\linewidth]{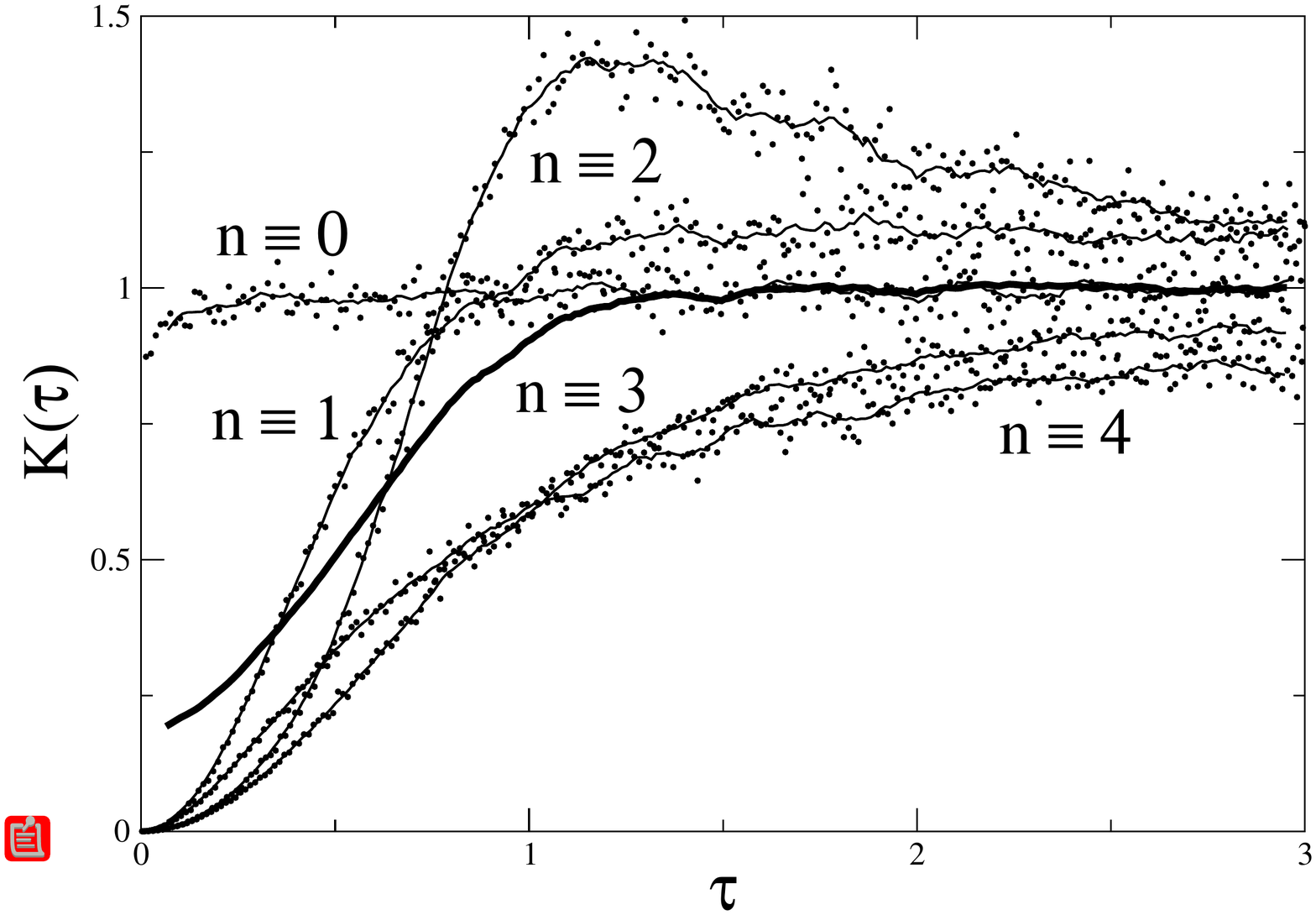}\\
(b)
\end{center}
\end{minipage}
\caption{Form factor for the matrix \eqref{map}  with  $\alpha=1/5$ and (a) $N=399\equiv -1\, \mathrm{mod}\, 5$
and (b) $N=398\equiv -2\, \mathrm{mod}\, 5$ averaged over $1000$ realisations. Points are values of $K\big (n/N\big )$ for integers $n$ with indicated residues modulo $5$.  Thin solid lines are guide for the eye.  
Thick solid lines indicate the average over all 5 residues:  $\frac{1}{5} \sum_{r=0}^4 K\big ((qt+r)/n\big )$. }
\label{map_02_399_398}
\end{figure}

For illustration,  the results of the direct calculation of the form factor for $\alpha=1/5$ and $N=399$ and $N=398$ are presented in figure~\ref{map_02_399_398}.  First,  eigenvalues of the matrix \eqref{map} were calculated numerically and then using  \eqref{K_trace} the form factor for different $n$ was computed. The result is averaged over $1000$ realisations of random phases. It is clear  that, indeed, for different residues of $n$ modulo $5$ the results are different and  when $n\equiv 0$ mod $5$ the form factor at small argument is close to 1 but for all other residues it starts at $0$.  The average over all 5 residues begins at $1/5$ as expected. 

Such clear picture appears when  the form factor  is calculated at  special values of $\tau$, $\tau=n/N$ with integer $n$. Computing it at arbitrary arguments leads to an irregular plot but, of course, the average curve remains unchanged.  

Exactly the same formulas can be applied for an irrational value of parameter $\alpha$. In this case one has
\begin{equation}
K(\tau)=\tau +2\, \tau \, \mathrm{Re} \left (\frac{e^{-\zeta  \tau }}{1-e^{-\zeta  \tau }}\right ),\qquad 
\zeta=1-e^{-2\pi i \alpha N}-2\pi i \alpha N 
\label{K_irrational}
\end{equation} 
The exponent $\zeta=2\sin^2(\pi \alpha N)+i \big [ \sin(2\pi \alpha N)-2\pi \alpha N\big ]$ has  a large imaginary part when $N\to\infty$. It means that the above expression is a strongly oscillated function of $\tau$. When averaged over a small interval of $\tau$ one obtains  $K(\tau)=\tau$ as it should be for the  ensemble of usual random matrices (GUE). This result follows without calculations from the fact that the average of all eigenvalues  $\Lambda_{\beta}^{\tau N}$ except  $\beta=0 $ equals  zero as a consequence of   rapidly changing   phases. (For even $N$ the term with $\beta=N/2$  is real but as it tends to zero at large $N$ its contribution is  negligible.) 

Notice that criterion \eqref{tanner_condition} for matrix~\eqref{map} with irrational $\alpha$ is not fulfilled. Nevertheless the spectral statistics of such matrix is close to GUE statistics. This example illustrates a new mechanism for the appearance of random matrix statistics.  The contribution of higher eigenvalues of the transition matrix \eqref{tanner_matrix} decreases not because a gap between the first and the second eigenvalues as has been proposed in \eqref{tanner_condition}  but due to rapid oscillations  for  large matrix dimensions.  

%=========================

\section{Symmetric barrier billiard with $\mathbf{h/b=1/2}$}\label{symmetric_BB}

%--------------------------------------------

The central problem  of the paper is the determination  of level compressibility for the $B$-matrices given by \eqref{B_matrix} and \eqref{L_m} by employing  the method proposed in \cite{tanner} and used in the previous Section for matrices derived from the quantisation of an interval-exchange map. The simplicity of treatment of interval-exchange matrices comes from the fact that their transition matrices are circulant matrices whose eigenvalues are known exactly. For the $B$-matrices calculations are more complicated as no explicit formulas for eigenvalues of the corresponding transition matrix.  
\begin{equation}
T_{m,n}=|B_{m,n}|^2=\frac{L_m^2\,L_n^2}{(x_m+x_n)^2}, \qquad m,n=1,\ldots,N
\label{T_BB}
\end{equation} 
are available. 

This section is devoted to the investigation of the $B$-matrix corresponding to the symmetric barrier billiard
with ratio $h/b=1/2$.  In this case $q=2$, $b-h=h$, and the second part of the vector $\vec{x}$ in \eqref{x_m_q} coincides  with the third one.  Now  trivial eigenfunctions can be  removed by considering a desymmetrised rectangular billiard  with height $h=b/2$ and imposing the Neumann boundary conditions for negative $x$ and $y=h$.  It is   equivalent of dropping the second part of vector \eqref{x_m_q} and taking  coordinates $x_m$ as follows \cite{billiard_1}
\begin{equation}
x_m=(-1)^{m+1}b\sqrt{k^2-\frac{\pi^2 m^2}{b^2}},\qquad m=1,\ldots,N,\qquad N=\left [ \frac{kb}{\pi}\right ] .
\label{x_symmetric}
\end{equation}
Odd (resp., even) indices describe the first (resp., the third) part of vector \eqref{x_m_q}.
 
\begin{figure}
\begin{minipage}{.49\linewidth}
\begin{center}
\includegraphics[width=.99\linewidth]{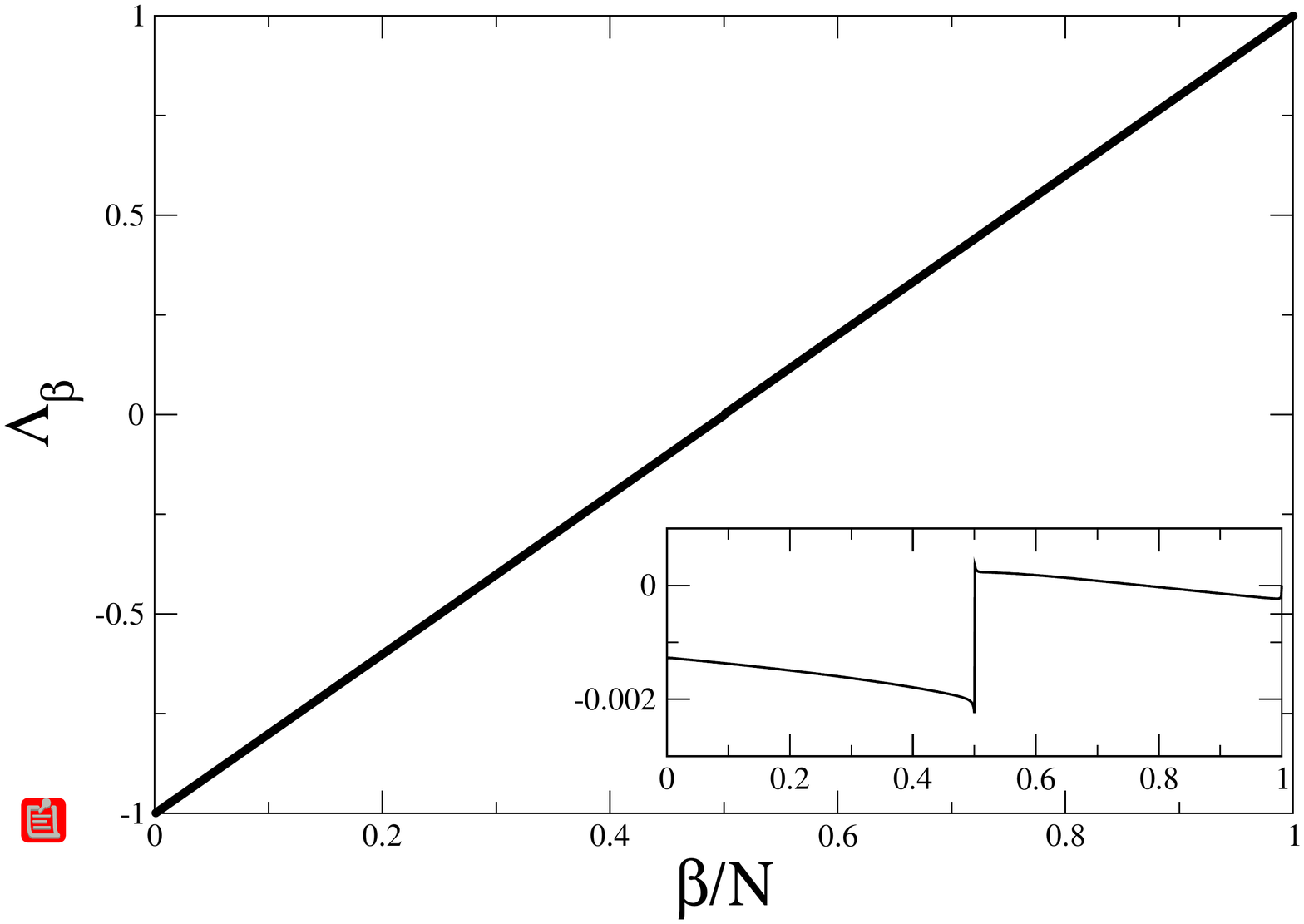}\\
(a)
\end{center}
\end{minipage}
\begin{minipage}{.49\linewidth}
\begin{center}
\includegraphics[width=.99\linewidth]{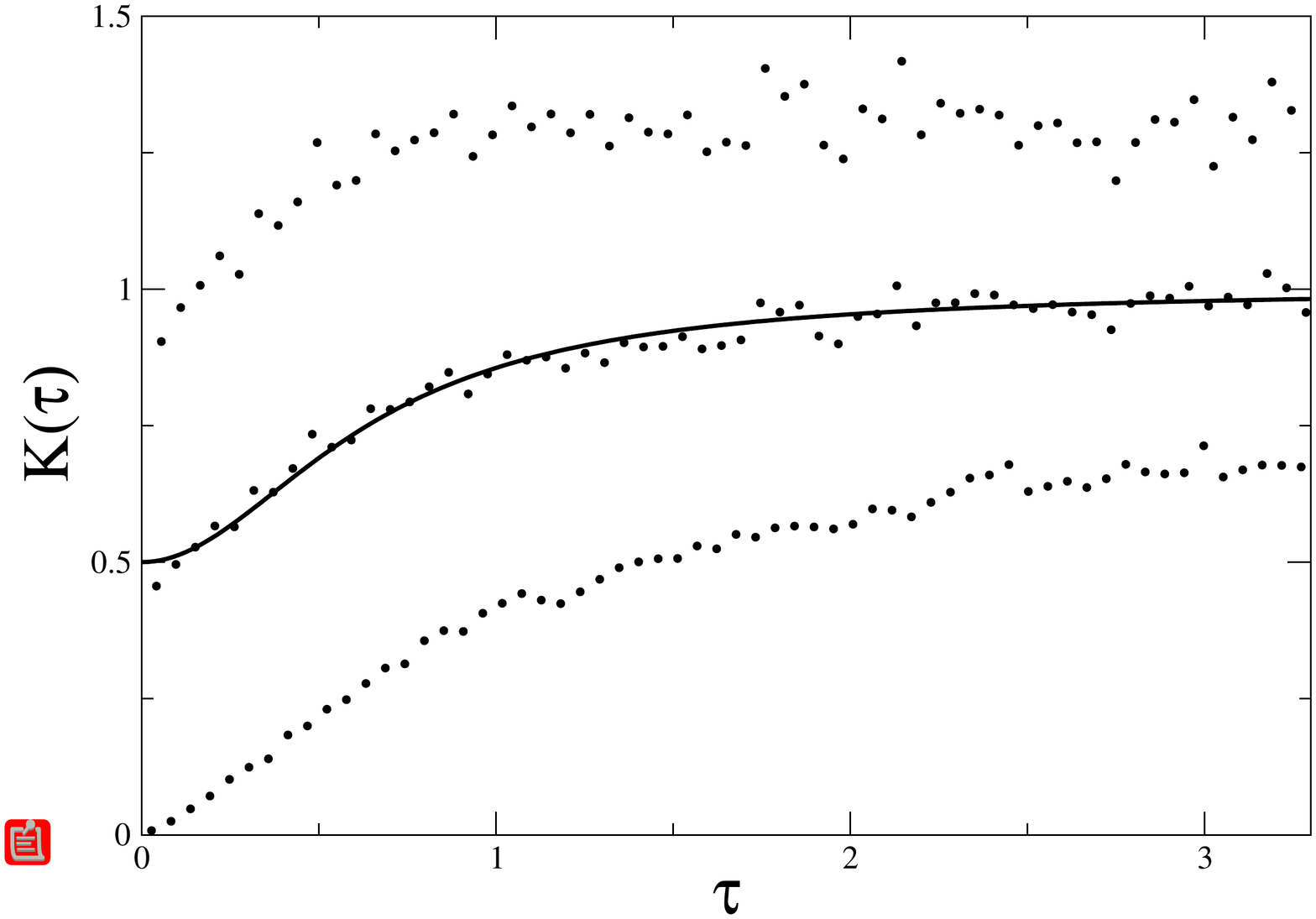}\\
(b)
\end{center}
\end{minipage}
\caption{(a) Transition matrix spectrum  for  symmetric barrier billiard with  $h/b=1/2$, $b=\pi$, $k=1000.5$, and   $N=1000$. Insert shows the difference between true eigenvalues $\Lambda_{\beta}$ and the straight line $2\beta/N-1$.  (b) Form factor for symmetric barrier billiard with $h/b=1/2$,   $b=\pi $, $k=400.5$, and $N=400$. Data   are averaged over 1000 realisations of random phases.  The upper dots are $K\big (2n/N\big )$, the lower dots are $K\big( (2n-1)/N\big )$, and the middle dots correspond to $\Big (K\big (2n/N\big )+K\big ((2n-1)/N\big)\Big )/2$. Solid line is the semi-Poisson prediction $K(\tau)=(2+\pi^2 \tau^2)/(4+\pi^2 \tau^2)$ \cite{gerland}.}
\label{transition_symmetric}
\end{figure}

The numerically calculated spectrum of the transition matrix in this case is presented in figure~\ref{transition_symmetric}(a).

To calculate this spectrum (or, at least, the behaviour of largest-moduli eigenvalues) analytically  a kind of paraxial approximation has been developed. It is based on  the fact that the main ingredient of matrices with intermediate statistics is a linear fall-off of matrix elements from the diagonal \cite{levitov, altshuler_levitov}. In the simplest setting it means that 
\begin{equation*}
M_{m,n}\sim \frac{R_{m,n}}{m-n+\mathrm{const.}} , \qquad m,n\gg 1, \qquad m-n=\mathcal{O}(1).  
\end{equation*}
Therefore it is natural to assume that the most important contributions come from the pole terms with  $R_{m,n}\approx R_{m,m}$. This type of approximation  can be done directly from the definition \eqref{L_m} as it is demonstrated  in  Appendix~\ref{direct_calculation}. According to these results the $T$-matrix in the paraxial approximation is  a block matrix
\begin{equation}
T =\left ( \begin{array}{cc} 0_{\mathrm{o,o}} &t_{\mathrm{o,e}} \\ t^{T}_{\mathrm{o,e}} & 0_{\mathrm{e,e}}  \end{array} \right ), \qquad t_{\mathrm{o,e}}\equiv  t_{2m-1,2n}=\frac{1}{\pi^2(n-m+1/2)^2}\, .  
\label{block_T}
\end{equation} 
Here subscripts  'o' and 'e' indicate odd and even indices respectively. 

It is instructive to get this answer without the knowledge of the exact $B$-matrix. One can achieved  it by using the instantaneous approximation used in quantum mechanics when the interaction changes suddenly.  In optics such approximation is analogous  to  the Fraunhofer diffraction. In the barrier billiard it  corresponds to the situation when a wave  with large momentum quickly moving  in a channel  enters into another channel (cf., figure~\ref{barrier_fig}(b)). In  the instantaneous approximation  eigenfunctions in the new channel  are just re-expansion of initial  eigenfunctions into a complete set of eigenfunctions with correct boundary conditions inside the final channel. 
   
Consider  a normalised wave with the Neumann boundary conditions at $y=h=b/2$ and the Dirichlet ones at $y=0$ 
\begin{equation*}
\psi^{(1)}_{2m-1}(x,y)=\frac{2}{\sqrt{b}} \sin \Big( \frac{ (2m-1)\pi }{b}y \Big)\exp\big (ip_{2m-1}^{(1)} x\big ) , \qquad x<0,\qquad 0\leq y \leq b/2
\end{equation*}
propagating in the desymmetrised  barrier billiard at negative $x$.   When it penetrates  into the region of positive $x$  it has to be expanded into correct waves propagating inside that region
\begin{equation*}
\psi^{(1)}_{2m-1}(x,y)=\sum_{n=1}^{\infty} S_{2m-1,2n} \psi^{(3)}_{2n}(x,y)
\end{equation*}
where $\psi^{(3)}_{2n}(x,y)$ are waves obeying the Dirichlet boundary conditions at $y=0$  and $y=h=b/2$
\begin{equation*}
\psi^{(3)}_{2n}(x,y)=\frac{2}{\sqrt{b}} \sin \Big( \frac{2 \pi n}{b}y \Big)\exp\big (ip_{n}^{(3)} x\big) , \qquad x>0,\qquad 0\leq y \leq b/2\,  .
\end{equation*}
 Coefficients $S_{2m-1,2n}$ are the $S$-matrix for this process. In the paraxial approximation they are calculated as follows (notice that in the paraxial approximation  $p_{2m-1}^{(1)}\approx p_{n}^{(3)}$)
\begin{equation*}
S_{2m-1,2n}^{(\mathrm{p.a.})}=\frac{4}{b} \int_{0}^{b/2}  \sin\Big( \frac{\pi (2m-1)}{b}\Big )\sin\Big( \frac{2\pi n}{b} y\Big)dy
= \frac{(-1)^{m+n}}{\pi^2(n-m+1/2)}+\frac{(-1)^{m+n}}{\pi^2(n+m-1/2)} \, .
\end{equation*}
Taking into account only the pole term (and symmetry of the $S$-matrix) one obtains for 
 the $T$-matrix exactly the same expression as  \eqref{T_limit}. 

Thus the transition matrix \eqref{block_T} is a block Toeplitz matrix. It is plain that its eigenvalues $\Lambda=\pm \sqrt{\lambda_{\beta}}$ where $\lambda_{\beta}$  are eigenvalues of  a $N_1\times N_1$ matrix (with $N_1=[N/2]$) 
\begin{equation*}
P_{m,n}=\sum_{k=1}^{N_1}  t_{m,k} t_{n,k},\qquad t_{m,k}=\frac{1}{\pi^2(k-m+1/2)^2} \, . 
\end{equation*}
Dominant contributions to the sum come from  regions $k\sim m$ and $k\sim n$. Due to a quick decrease of the summands   the finite summation over  $k$  can safely  be substitute in the limit $N_1\to\infty$ by  the sum over all integer $k$ 
\begin{equation*}
P_{m,n}\approx \sum_{k=-\infty}^{\infty}  t_{m,k} t_{n,k}\, . 
\end{equation*}
Using \eqref{standard_formulas} the necessary sum is easily calculated and the result is 
\begin{equation}
P_{m,n}=t_{m-n}, \qquad t_0=\frac{1}{3},\qquad t_{r\neq 0}= \frac{2}{\pi^2 r^2}\, . 
 \label{P_matrix}
  \end{equation}
 This formula is valid when $m,n\gg 1$ and $m-n=\mathcal{O}(1)$. 
 
 Matrix~\eqref{P_matrix} is a Toeplitz matrix  with quickly decreasing matrix elements. It is well known that eigenvalues of $N\times N$ Toeplitz matrix can be asymptotically calculated as follows  (see, e.g.,\cite{szego}-\cite{rambour} and references therein) 
 \begin{equation}
 \lambda_{\beta}=f\left ( \frac{\beta}{2 N}+\mathcal{O}\left (\frac{1}{N}\right ) \right ) ,\qquad  \beta=1,\ldots,N 
 \label{Toeplitz_eigenvalues} 
 \end{equation} 
 where function $f(x)$ called the symbol is the Fourier series of   $t_r$
 \begin{equation*}
 f(x)= \sum_{r=-\infty}^{\infty} t_r e^{2\pi i r x}.
 \end{equation*}
 (More precise formulas can be found  in the above references.)  
 
 Using~\eqref{Bernoulli} and \eqref{B_2_3} one finds that  the symbol of  matrix~\eqref{P_matrix}  is 
 \begin{equation*}
 f(x)=\frac{1}{3}+\frac{2}{\pi^2}\sideset{}{'}\sum_{r=-\infty}^{\infty}  \frac{e^{2\pi i r x}}{r^2} =\frac{1}{3}+4B_2(x)=(1-2x )^2. 
 \end{equation*}
 Therefore eigenvalues of the $P$-matrix  for large $N_1$ are 
 \begin{equation*}
 \lambda_{\beta}\approx \left (1-\frac{\beta}{N_1}\right )^2, \qquad \beta=1,\ldots,N_1\, .
 \end{equation*}
Eigenvalues of block matrix \eqref{block_T} $\Lambda=\pm \sqrt{\lambda_{\beta}}$.  Taking into account that the dimension of matrix \eqref{block_T} is $N\approx 2N_1$ one concludes that approximately its eigenvalues  are\begin{equation}
\Lambda=\pm \left (1-\frac{2\beta}{N} \right ),\qquad   \beta =1,\ldots,\tfrac{1}{2}N
\label{approx_Lambda}
\end{equation} 
With the corresponding redefinition  of index $\beta$  these eigenvalues can be rewritten in the form 
\begin{equation*}
\Lambda_{\beta}\approx \frac{2 \beta}{N}-1,\qquad \beta =1,\ldots,N
\end{equation*}
which agrees well with numerical calculations (see figure~\ref{transition_symmetric}(a)).  

The form factor in the diagonal approximation is related with transition matrix eigenvalues by \eqref{K_lambda}
\begin{equation*}
K^{\mathrm{(diag)}}\left (\frac{n}{N}\right )=\frac{n}{N}\sum_{\beta}\Lambda_{\beta}^n \, .
\end{equation*}
As $\Lambda_{N-\beta}=-\Lambda_{\beta}$ for $\beta =1,\ldots,N-1$ (which is  a consequence of the block structure of the transition matrix \eqref{block_T}) the form factor $K^{\mathrm{diag}}(n/N)$ with odd $n$ in the diagonal approximation tends to  zero when $\tau=n/N\to 0$
\begin{equation}
K^{(\mathrm{diag})}(\tau) \underset{\tau\to 0}{\longrightarrow} 0\, .
\label{K_odd}
\end{equation} 
But for even $n$ one gets a different answer. Eq.~\eqref{approx_Lambda}  may  not be accurate for extreme eigenvalues with small $\beta$.   For $\tau=2n/N$ one can separate contribution of small $\beta<\beta_0$ and the rest for which \eqref{approx_Lambda} is a good approximation
\begin{equation}
K^{(\mathrm{diag})}(\tau)=\tau \left (\mathrm{const} +2\sum_{\beta=\beta_0}^{N/2} \left (1-\frac{2\beta }{N}\right )^{\tau N} \right )\underset{N\to\infty}{\longrightarrow}\tau \left (\mathrm{const}+\frac{2e^{-2 j_0\tau}}{1-e^{-2\tau}} \right )\underset{\tau\to 0}{\longrightarrow}1\, . 
\label{K_even}
\end{equation}
As has been discussed in the previous Section it means that the spectral compressibility of the $B$-matrix for symmetric barrier billiard coincides with the semi-Poisson value
\begin{equation}
\chi\equiv K(0)=\frac{1}{2}\, .
\label{K_1_2}
\end{equation}
For illustration, the form factor for the symmetric billiard  calculated numerically by direct diagonalisation of $400\times 400$  matrices  \eqref{B_matrix} with coordinates given by  \eqref{x_symmetric} and  averaged over $1000$ realisations is shown in figure~\ref{transition_symmetric}(b).  Two branches corresponding to odd and even $n$ are clearly seen. The average over odd and even values agrees well with the semi-Poisson expression for the form factor \cite{gerland} and,  in fortiori, the level compressibility is $1/2$ as in \eqref{K_1_2}.  

%=======================
\section{Barrier billiard with irrational ratio $\mathbf{h/b}$}\label{irrational_BB}
%----------------------------------------

\begin{figure}
\begin{minipage}{.49\linewidth}
\begin{center}
\includegraphics[width=.99\linewidth]{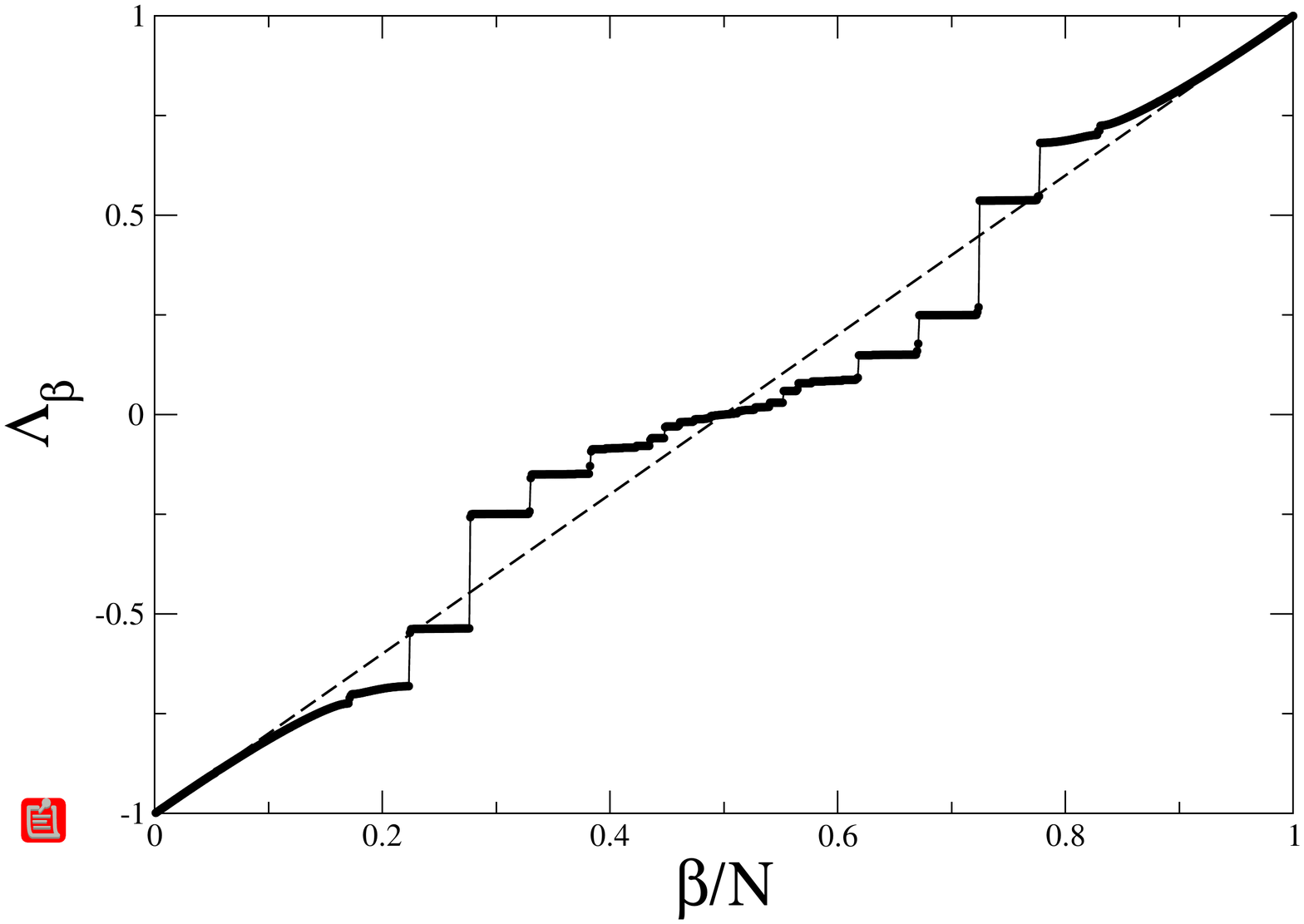}\\
(a)
\end{center}
\end{minipage}
\begin{minipage}{.49\linewidth}
\begin{center}
\includegraphics[width=.99\linewidth]{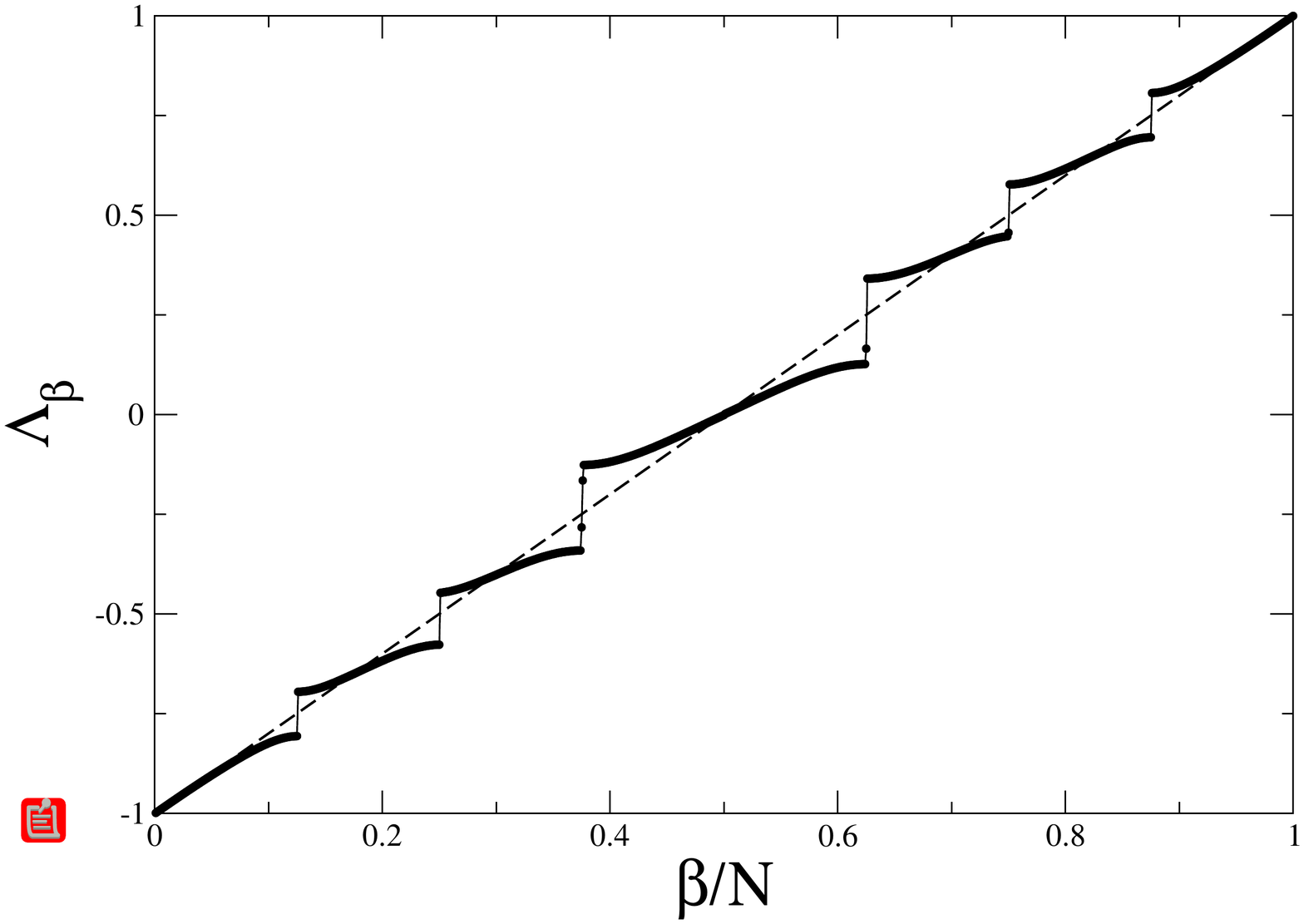}\\
(b)
\end{center}
\end{minipage}
\caption{Spectra of the transition matrix for barrier billiard with (a) $h/b=1/\sqrt{5}$, $b=\pi$, $k=500.5$, and  $N=999$ (cf. \eqref{N_irrational}) (b)  $h/b=2/5$, $b=\pi$, $k=650.5$,and $N=1040$ (cf. \eqref{N_m_q}). Straight dashed line in the both figures is  $\Lambda_{\beta}=2\beta/N-1$.  } 
\label{transition_eigenvalues}
\end{figure}

 The  transition matrices for general barrier billiard with  off-centre barrier  remain the same as in \eqref{T_BB} but coordinates $x_m$ should have the form \eqref{x_irrational} for irrational ratio $h/b$ and \eqref{x_m_q} for rational $h/b=m/q$. The direct calculations of eigenvalues of these matrices reveal that they  are more complicated that the ones for symmetric billiard with $h/b=1/2$ discussed in the previous Section. As an example, in figure~\ref{transition_eigenvalues}  the spectra of the transition matrices with $h/b=1/\sqrt{5}$  and $h/b=2/5$ are presented. It is clearly seen that, though eigenvalues with small moduli are quite irregular and have gaps, largest moduli eigenvalues are well described by a straight line $\Lambda_{\beta}=2\beta/N-1$.

This Section is concentrated on the analytical treatment  of billiards with irrational ratio $h/b$. As in the previous Section the first step consists in the calculation  of paraxial $S$-matrix for the scattering inside the slab with a barrier as in figure~\ref{barrier_fig}(b). It can  easily be done  in the instantaneous approximation exactly as above. In such approximation only  transitions  from channel $1$ to $2$ and to $3$ and  their  inverse are non-zero.  One has 
\begin{eqnarray*}
S_{m,n}^{1\to 3}&=&\frac{2}{\sqrt{bh}} \int_0^{h} \sin\Big (\frac{\pi m}{b}y \Big)\sin\Big (\frac{\pi n}{h}y \Big)dy\nonumber\\
&=&\frac{(-1)^n \sin (\pi h m/b)}{\pi \sqrt{bh}}\left ( (m/b-n/h)^{-1}-(m/b+n/h)^{-1} \right ).
\end{eqnarray*}
Similarly
\begin{eqnarray*}
S_{m,n}^{1\to 2}&=&\frac{2}{\sqrt{b(b-h) }} \int_{h}^{b} \sin\Big (\frac{\pi m}{b}y \Big)\sin\Big (\frac{\pi n}{b-h}(b-y) \Big)dy\nonumber\\
&=&\frac{(-1)^n \sin (\pi m  h/b) }{\pi \sqrt{b(b-h)}} \left ((m/b-n/(b-h))^{-1}-(m/b+n/(b-h))^{-1} \right )\, .
\end{eqnarray*}
The transition matrix $T=|S|^2$ also has the same  block structure. Retaining only the pole (the first) terms (and slightly changing the notations)  one obtains 
\begin{equation}
T =\left ( \begin{array}{ccc} 0_{1,1} &t^{1\to 2}_{n_1,n_2} & t^{1\to 3} _{n_1,n_3} \\ t^{1\to 2}_{n_1,n_2} & 0_{2,2}& 0_{2,3}  \\
t^{1\to 3}_{n_1,n_3} & 0_{3,2} &0_{3,3}
 \end{array} \right )
 \label{block_general}
\end{equation}
where $n_i=1,\ldots,N_i$ with $N_i$ given by \eqref{N_irrational} and  
\begin{equation}
t^{1\to 2}_{n_1,n_2}=\frac{ z_2 \sin^2 (\pi n_1  h/b)}{\pi^2 \big ( z_2 n_1-n_2 \big )^2 }, \qquad 
t^{1\to 3}_{n_1,n_3}=\frac{ z \sin^2 (\pi n_1  h/b)}{\pi^2  \big (z n_1 -n_3 \big )^2 },\qquad  z_2=1-z,\qquad  z=\frac{h}{b}.
\label{t_paraxial}
\end{equation}
Due to the block structure of the transition matrix \eqref{block_general} it follows that its eigenvalues $\Lambda$ are determined  by the relation  $\Lambda^2=\lambda_{\beta}$ where $\lambda_{\beta}$ are eigenvalues of $N_1\times N_1$ matrix
\begin{equation*}
P_{m,n}=\sum_{k=1}^{N_2} t^{1\to 2}_{m,k}t^{1\to 2}_{n,k}+\sum_{k=1}^{N_3} t^{1\to 3}_{m,k}t^{1\to 3}_{n,k}
\end{equation*}
For large matrix dimension  the summation can be extended over all integer $k$ and the sums can be calculated explicitly by using \eqref{useful_formulas}  from Appendix~\ref{divers}. The results are 
\begin{equation}
P_{m,m}=\frac{z^2+(1-z)^2}{3} (3-2\sin^2(\pi m z)),
\label{diagonal_P}
\end{equation}
and for $m\neq n$
\begin{equation}
 P_{m,n}=\frac{2\big (\sin^2(\pi z m)+\sin^2(\pi n z)\big ) }{\pi^2(m-n)^2}-\frac{2 (1-2z)\sin(\pi m z) \sin(\pi nz) \sin(\pi (m-n)z)}{\pi^3 z(1-z) (m-n)^3}\, . 
 \label{off_diagonal_P}
\end{equation}
This matrix is a combination of Toeplitz terms depended on the difference $m-n$ and oscillating terms (which explains the existence of forbidden zones in its spectrum, see figure~\ref{transition_eigenvalues}(a). 

Due to the unitarity of the $B$-matrix the exact transition matrix $T=|B|^2$ has the largest eigenvalue equals $1$ whose  corresponding eigenvector is $(1,1,\ldots,1)$. It is natural (and is confirmed by calculations)  that eigenvectors of the $P$-matrix  corresponding  to large moduli eigenvalues are slowly  varying functions. Consequently,  all oscillating terms in \eqref{diagonal_P} and \eqref{off_diagonal_P} for large $m$ and $n$ could  be ignored. 
These arguments lead to the following recipe of the next step of approximation. Put $m=n+r$ and average all matrix elements of the $P$-matrix over quickly changing phase $\pi n z$. The calculations are straightforward and 
\begin{equation}
\langle {P}_{m,n}\rangle\equiv \lim_{N\to\infty} \frac{1}{N} \sum_{n=1}^{N}P_{n+r,n}=f_{m-n}
\label{average_P_irrational}
\end{equation}
where
\begin{equation*}
 f_0=\frac{2(z^2+(1-z)^2)}{3}, \qquad f_{r\neq 0}=\frac{2}{\pi^2r^2}-\frac{(1-2z)\sin(2\pi r z)}{2\pi^3 z(1-z) r^3}\,  .
\end{equation*}
Eigenvalues of such matrix for large $N$ are calculated by the Fourier transform of this symbol
\begin{equation*}
\lambda_{\beta}=\sum_{r=-\infty}^{\infty} f_r e^{2\pi i r x}, \qquad x=\frac{\beta}{2 N_1}\, .
\end{equation*} 
The necessary sums are expressed through the Bernoulli polynomials \eqref{Bernoulli}, \eqref{B_2_3} and the result is 
\begin{equation*}
\lambda_{\beta}=\frac{2}{3}(z^2+(1-z)^2)+4B_2(x)-\frac{1-2z}{3 z(1-z)}\Big ( B_3(\{x+z\})-B_3(\{x-z\}) \Big ).  
\end{equation*}
From the beginning one can assume  that $h<h-b$, i.e.,  $z=h/b<1/2$ (the case $h/b=1/2$ was discussed in Section~\ref{symmetric_BB} ). Then   
\begin{equation}
\lambda_{\beta}=\left \{ \begin{array}{cc}   1-4x+ \dfrac{x^2}{z(1-z)} , & 0\leq x \leq z\\
\dfrac{1-2z+2z^2}{1-z}-\dfrac{2x(1-x)}{1-z}, & z\leq x\leq \frac{1}{2} \end{array}\right . ,\qquad x=\frac{\beta}{2N_1}\, .
\label{approximative}
\end{equation}
As eigenvalues of the block matrix \eqref{block_general}  $\Lambda=\pm \sqrt{\lambda_{\beta}}$ it follows that close to maximum value (i.e., with small $\beta$) 
\begin{equation*}
\Lambda\approx \pm \left (1-\frac{\beta}{N_1}+\mathcal{O}\left (\frac{1}{N_1}\right )\right )\approx \pm \left (1-\frac{2\beta}{N}\right )
\end{equation*}
 As in the calculation of the form factor \eqref{K_lambda} small moduli eigenvalues are irrelevant one can ignore 
higher order terms in the above expression which gives the same expression as in \eqref{approx_Lambda}. It means that the level compressibility of barrier billiards with irrational ratio $h/b$ has the same value as in the preceding Sections $\chi=\tfrac{1}{2}$.

\begin{figure}
\begin{minipage}{.49\linewidth}
\begin{center}
\includegraphics[width=.99\linewidth]{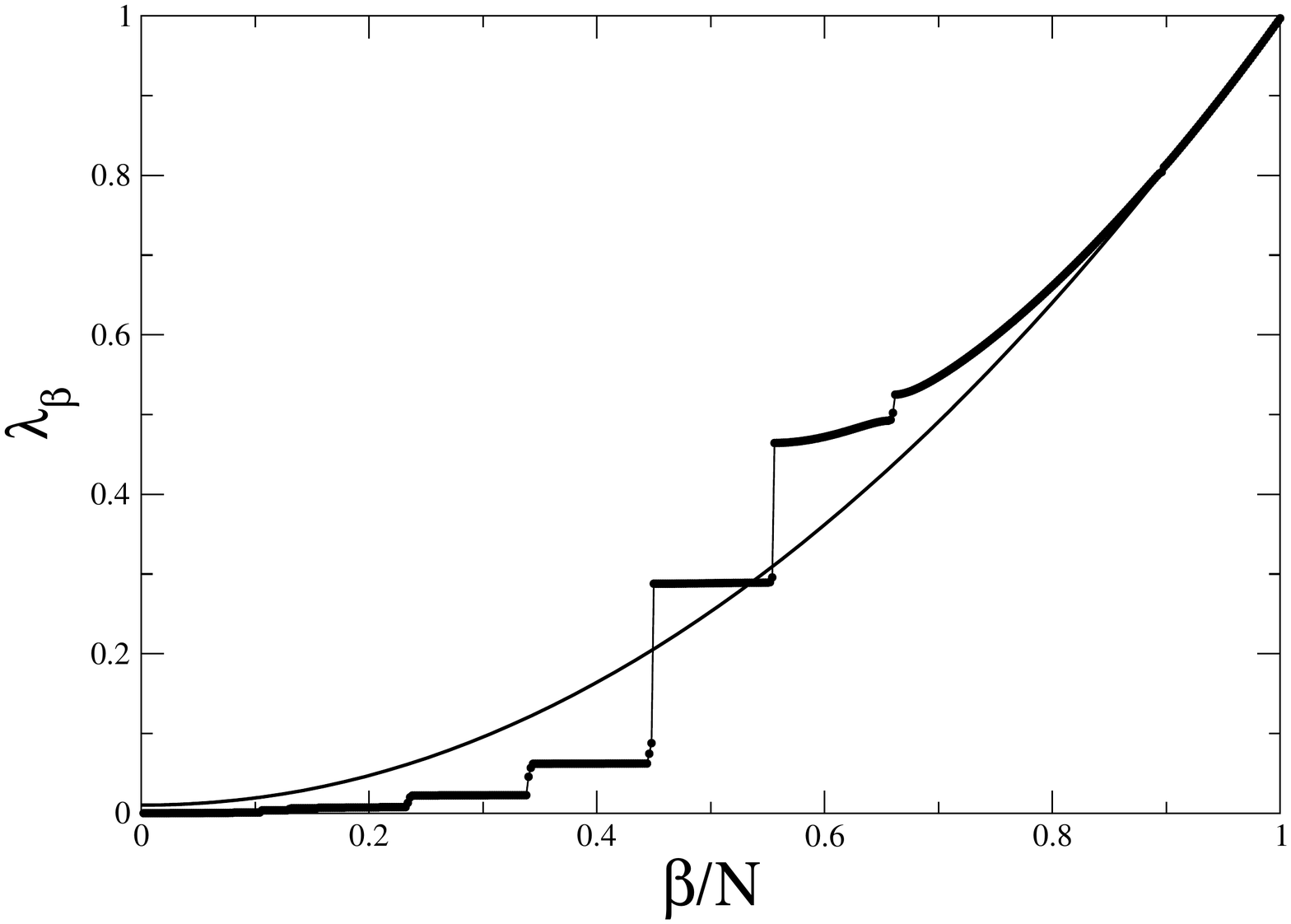}\\
(a)
\end{center}
\end{minipage}
\begin{minipage}{.49\linewidth}
\begin{center}
\includegraphics[width=.99\linewidth]{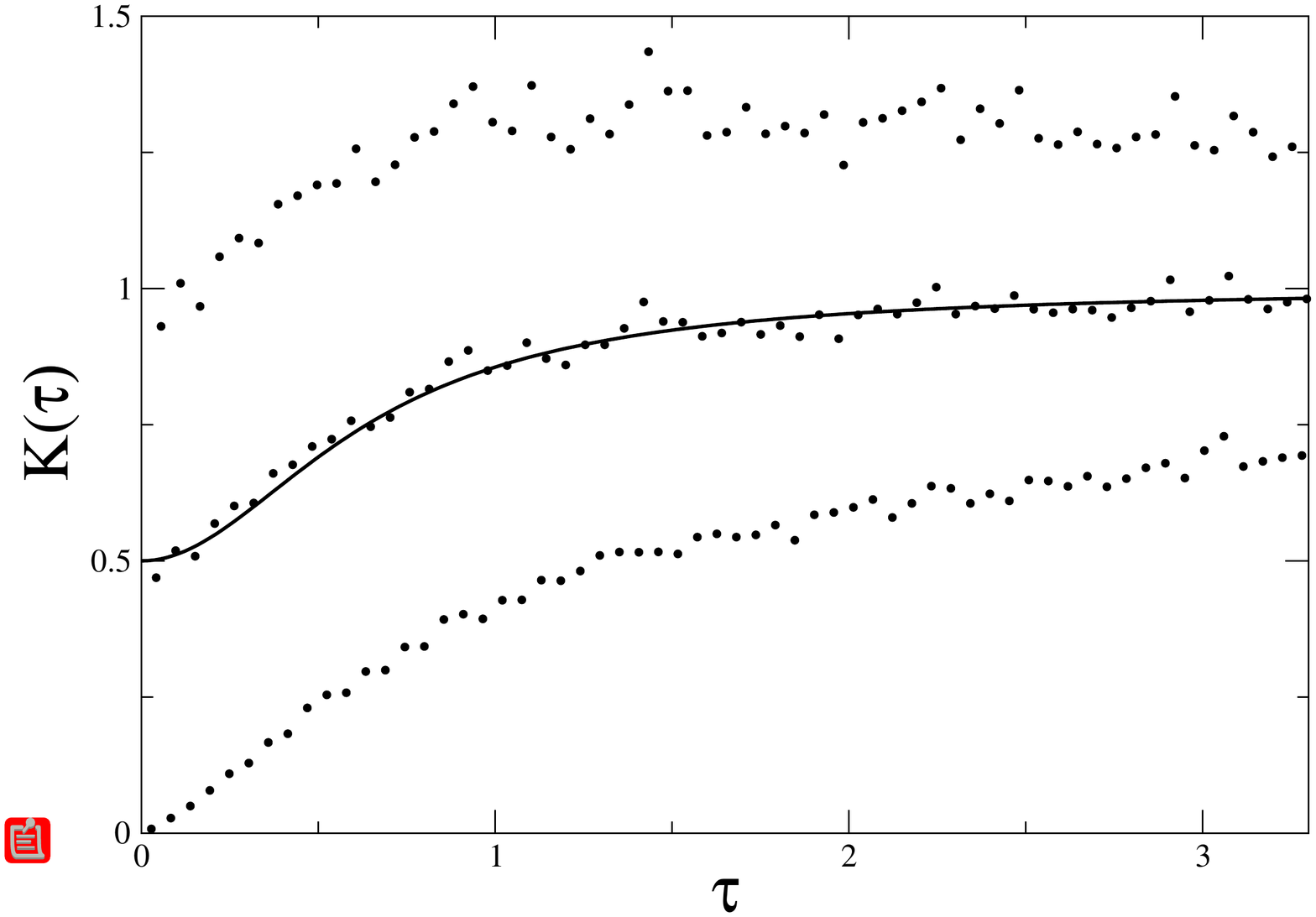}\\
(b)
\end{center}
\end{minipage}
\caption{(a) Eigenvalues of the $P$-matrix \eqref{diagonal_P}, \eqref{off_diagonal_P} for for $h/b=1/\sqrt{5}$ for the same parameters as in figure~\ref{transition_symmetric}(a). Solid line indicates  approximate expressions \eqref{approximative}. (b) Form factor for $h/b=1/\sqrt{5}$, $b=\pi$, $k=200.5$, $N=399$ averaged over 1000 realisations. Other notations are  as in figure~\ref{transition_symmetric}(b).} 
\label{eigenvalues_1_sqrt_5}
\end{figure}

In figure~\ref{eigenvalues_1_sqrt_5}(a) the above formulas are compared with the results of direct calculations for the $P$-matrix with $h/b=1/\sqrt{5}$. As has been demonstrated, approximate expression \eqref{approximative} is tangent to the exact spectrum close to $1$. The form factor computed numerically for the same ratio $h/b$ is presented in figure~\ref{P_matrix}(b). The agreement with the above result is clearly seen. 

%=======================
\section{Barrier billiard with rational ratio $\mathbf{h/b=p/q}$}\label{rational_BB} 
%----------------------------------------

The calculation of transition matrix eigenvalues when the ratio $h/b$ is a rational number can be done by a  similar method. An additional difficulty in such case is that one has to select special combinations of states in the second and the third channels to remove trivial eigenvalues equal zero on the whole line passing through the barrier. It has been discussed in detail in \cite{billiard_2} and briefly reviewed in Appendix~\ref{resonance}.  Combining all terms together one concludes that  the transition matrix when $h/b=p/q$ with $p$ and $q$ being co-prime integers has the block form similar to \eqref{block_general} but with one more block
\begin{equation}
T =\left ( \begin{array}{cccc} 0_{1,1} &t^{1\to 2}_{n_1,n_2} & t^{1\to 3} _{n_1,n_3}&t^{1\to 4}_{n_1, n4}  \\ t^{1\to 2}_{n_1,n_2} & 0_{2,2}& 0_{2,3} &0_{2,4}  \\
t^{1\to 3}_{n_1,n_3} & 0_{3,2} &0_{3,3}& 0_{3,4}\\
t^{1\to 4}_{n_1,n_4}& 0_{4,2} & 0_{4,3}& 0_{4,4}
 \end{array} \right ) .
 \label{block_rational}
\end{equation}
Here indices $n_j=1,\ldots N_j^{\mathrm{(r)} }$ have the following restrictions
\begin{equation*}
n_1\not\equiv 0 \mod q,\qquad n_2\not\equiv  0 \mod q-p,\qquad n_3\not\equiv 0 \mod p  
\end{equation*}
and 
\begin{equation}
N_1^{\mathrm{(r)}}=N_1-N_0,\qquad N_2^{\mathrm{(r)}}=N_2-N_0,\qquad N_3^{\mathrm{(r)}}=N_3-N_0,\qquad N_4^{\mathrm{(r)}}=N_0
\label{N_resonant}
\end{equation}
with $N_1,N_2,N_3$ given by \eqref{N_irrational} and $N_0$ is determined by \eqref{N_0} or \eqref{N_m_q}. The total matrix dimension is $N^{(\mathrm{r})}=\sum_{j=1}^{4}N_j^{\mathrm{(r)}} = N_1+N_2+N_3-2N_0$ as in \eqref{N_m_q}. 

Matrices $t^{1\to 2}$ and $t^{1\to 3}$ are the same as in \eqref{t_paraxial} and  $t^{1\to 4}$ given by  \eqref{t_2_3}  from  Appendix~\ref{resonance}
\begin{equation*}
t_{n_1,n_4}^{1\to 4} =\frac{\sin^2(\pi m h/b)}{\pi^2p (q-p) (n_1/q-n_4)^{2}}\, .
\end{equation*}
The eigenvalues of block matrix \eqref{block_rational} $\Lambda$ are $\Lambda=\lambda_{\beta}^2$ where $\lambda_{\beta}$ with $\beta=1,\ldots,N_1$ are eigenvalues  of matrix (superscript (res) indicates that the matrix describes  the resonance case $h/b=p/q$)
\begin{equation*}
P_{m,n}^{\mathrm{(res)}}=\sum_{k\neq 0 \mod (q-p) }t_{m,k}^{1\to 2} t_{n,k}^{1\to 2}+\sum_{k\neq 0 \mod p }t_{m,k}^{1\to 3} t_{n,k}^{1\to 3}+\sum_{k }t_{m,k}^{1\to 4} t_{n,k}^{1\to 4}\, . 
\end{equation*} 
Using an evident relation 
\begin{equation*}
\sum_{k\neq 0\mod r}f(k)=\sum_k f(k)-\sum_{k}f(rk) 
\end{equation*}
and \eqref{useful_formulas} the above sums can be explicitly calculated. 

The results are 
\begin{equation}
P_{m,m}^{\mathrm{(res)}}=\frac{1}{3q^2}\big (p^2 +(q-p)^2\big )\big ( 3-2\sin^2(\pi pm/q) \big )+\frac{2\sin^4(\pi pm/q) \big (3-2\sin^2(\pi m/q)\big )}{3p(q-p)q^2\sin^4(\pi m/q)}
\label{P_diagonal_resonant}
\end{equation}
and when $m\neq n$
\begin{equation}
P_{m,n}^{\mathrm{(res)}}=\frac{b_{m,n}}{(m-n)^2} -\frac{c_{m,n}\sin(\pi p (m-n)/q)}{(m-n)^3}
-\frac{d_{m,n}\sin(\pi (m-n)/q)}{(m-n)^3}
\label{P_off_diagonal_resonant}
\end{equation}
where
\begin{eqnarray}
b_{m,n}&=&\frac{2}{\pi^2} \big ( \sin^2(\pi p m/q)+\sin^2(\pi  p n /q) \big )\nonumber\\
&+&\frac{2 \sin^2(\pi pm/q)\sin^2(\pi  p n /q) }{p(q-p)\pi ^2}\Big (\frac{1}{\sin^2(\pi m/q)}+\frac{1}{\sin^2(\pi n/q)}\Big)\label{resonant_coefficients} \\
c_{m,n}&=&\frac{2 q (q-2p)}{\pi^3 p(q-p)}\sin(\pi p m /q)\sin(\pi p n /q),\qquad 
d_{m,n}=\frac{4 q \sin^2 (\pi p m /q)\sin^2(\pi p n /q)}{\pi^3 p(q-p) \sin(\pi m/q) \sin(\pi n/q)}
\, . \nonumber
\end{eqnarray}
Though these expressions are indexed by integers $m$ and $n$ this notation is symbolic. The point  is that by construction these integers cannot be arbitrary but have to be not divisible by $q$. Let us ordered such numbers and let  $\nu(k)$ with $k=1,2,\ldots, $ be the $k^{\mathrm{th}}$ integer $\not \equiv 0 \mod q$. Then indices of matrix $P_{m,n}^{\mathrm{(res)}}$ have to be considered as follows: $m=\nu(j)$, $n=\nu(k)$ with  $j,k=1,2,\ldots, N_1^{\mathrm{(r)}}$ with $N_1^{\mathrm{(r)}}$ defined in \eqref{N_resonant}. In such notation matrix $P^{\mathrm{(res)}}$  is $N_1^{\mathrm{(r)}}\times N_1^{\mathrm{(r)}}$ matrix 
\begin{equation*}
P_{m,n}^{\mathrm{(res)}} \equiv P_{\nu(j),\nu(k)}^{\mathrm{(res)}}, \qquad j,k=1,\ldots,N_1^{\mathrm{(r)}}. 
\end{equation*}
The next step, as in the previous Section (cf., \eqref{average_P_irrational}),  consists in the calculation instead of the above exact expressions their mean values with fixed difference between the indices
\begin{equation*}
\langle P_{m,n}^{\mathrm{(res)}}\rangle(r) =\lim_{N\to\infty} \frac{1}{N}\sum_{n=1}^N P_{n+r,n}^{\mathrm{(res)}}
\end{equation*}  
where the both  integers $n$ and $n+r$ have to be not divisible by $q$. 

According to \eqref{P_diagonal_resonant} and  \eqref{P_off_diagonal_resonant} the  $P^{\mathrm{(res)}}$ matrix is a mixture of functions depending explicitly on the differences of indices and certain coefficients depending on indices modulus $q$. Only the latter requires 
the explicit averaging. Using \eqref{F_1}-\eqref{F_4} from  Appendix~\ref{divers} one obtains that 
\begin{eqnarray*}
& & \sum_{m=1}^{q-1}P_{m,m}=\frac{(p^2+(q-p)^2)(2q-3)}{3q^2} +\frac{2p^2(2q-3p)}{3(q-p)q^2},\\
& & \sideset{}{'}\sum_{m=1}^{q-1}b_{m,m+r}=\frac{2q(q-p+1-2\sin^2(\pi p r/q))}{\pi^2(q-p)},\\
& & \sideset{}{'}\sum_{m=1}^{q-1} c_{m,m+r}=\frac{q^2(q-2p) \cos(\pi p r/q)}{\pi^3 p (q-p)},\qquad  
\sideset{}{'}\sum_{m=1}^{q-1} d_{m,m+r}=\frac{q^2 \sin(2\pi p r/q)}{\pi^3 p (q-p) \sin(\pi r/q)}\, .
\end{eqnarray*}
Here it is taken into account that $p/q<1/2$. The superscript $\prime$ in these sums  indicates that the term with $m+r\equiv 0 \mod q$ is omitted. The latter condition implies that the  number of independent terms equal $q-1$ if $r\equiv 0 \mod q$ or $q-2$ otherwise.  Finally one obtains 
\begin{equation}
\langle P_{m,n}^{\mathrm{(res)}}\rangle =\alpha_1 \delta_{m,n}+f_{m-n}
\end{equation}
with
\begin{equation}
f_{m-n}=\left \{ \begin{array}{cc} \dfrac{\alpha_2}{(m-n)^2} , & (m-n)\equiv 0 \mod q \\  & \\
\dfrac{\alpha_3}{(m-n)^2}+\dfrac{\alpha_4 \cos(2\pi p r/q)}{(m-n)^2}+ \dfrac{\alpha_5 \sin(2\pi p r/q) }{(m-n)^3} , &(m-n)\equiv r \neq 0 \mod q 
\end{array}\right .
\end{equation}
where constants $\alpha_j$  are 
\begin{eqnarray}
\alpha_1&=& \frac{(p^2+(q-p)^2)(2q-3)}{3q^2(q-1)} +\frac{2p^2(2q-3p)}{3(q-p)q^2(q-1)}\qquad 
\alpha_2=\frac{2q(q-p+1)}{\pi^2 (q-p)(q-1)},\nonumber \\ 
\alpha_3&=&\frac{2q}{\pi^2 (q-2)},\qquad 
\alpha_4=\frac{2q}{\pi^2(q-p) (q-2)},\qquad 
\alpha_5 =\frac{q^2(2p-q-2)}{2\pi^3 p(q-p)(q-2))}\, .
\end{eqnarray}
Though this matrix depends only on the difference of indices $m-n$ it is not a Toeplitz matrix as $m$ and $n$ are not arbitrary numbers but only  integers  not divisible by $q$. Nevertheless one can argue that largest eigenvalues for large matrix dimension can be calculated by  a formula similar to Toeplitz matrices (which is a kind of variational method) 
\begin{equation}
\lambda_{\beta} =\frac{1}{N_1^{\mathrm{(r)}}}\sum_{j,k=1}^{N_1^{\mathrm{(r)}}} \langle P_{\nu(j),\nu(k)}^{\mathrm{(res)}}\rangle e^{2\pi i x(j-k)},\qquad x=\frac{\beta}{2 N_1^{\mathrm{(r)}}}\, .
\label{variational_principle}
\end{equation}
Here, as above, $\nu(k)$ is the $k^{\mathrm{th}}$ integer $\not \equiv 0 \mod q$.

In Appendix~\ref{divers} (see  \eqref{sum_S})  it is shown that such sum can be written as follows
\begin{eqnarray*}
\lambda_{\beta}&=&\alpha_1 + \alpha_2  \sum_{u=-\infty}^{\infty} \frac{e^{2\pi i x(q-1)u}}{(qu)^2}
+2 \mathrm{Re}\, \sum_{t=1}^{q-1} \left (1-\frac{t}{q-1}\right )\times \nonumber\\ 
&\times&  \sum_{u=-\infty}^{\infty} e^{2\pi ix((q-1)u+t)} \left (\frac{\alpha_3}{(u q+t)^2 } +
\frac{\alpha_4 \cos(2\pi p t/q)}{(u q+t)^2}+\frac{ \alpha_5 \sin(2\pi p r/q) }{(uq+t)^3}\right )\, .
\end{eqnarray*}
The first sum is calculated through the Bernoulli polynomial $B_2(x)$ (see \eqref{Bernoulli}). The last sums are expressed through two functions 
\begin{equation*}
G(x,r)=\sum_{k=-\infty}^{\infty}\frac{e^{2 \pi  i x (kq+r)}}{(kq+r)^2},\qquad F(x,r)=\sum_{k=-\infty}^{\infty}\frac{e^{2 \pi  i x (kq+r)}}{(kq+r)^3}\, .
\end{equation*}
The explicit expressions of these function can be obtained as follows. 

Define one more  function 
\begin{equation*}
g(x,r)=\sum_{k=-\infty}^{\infty}\frac{e^{2 \pi i x (kq+r)}}{kq+r}\, .
\end{equation*}
By the differentiation over  $x$ one has $G^{\prime}(x,r)=2\pi i g(x,r)$ and  $F^{\prime}(x,r)=2\pi i G(x,r)$. As the differentiation of $g(x,r)$ over $x$ gives the sum of $\delta$-function it is plain that $g(x,r)$ is the piece-wise constant function in interval $[j/q, (j+1)/q]$. Using \eqref{standard_formulas} one gets
\begin{equation*}
 g(j/q+x,r)=\frac{\pi}{q\sin(\pi r/q)}e^{i\pi r(2j+1)/q}\, .
\end{equation*}
Correspondingly,  function $G(x,r)$ is a piece-wise linear function in the same intervals
\begin{equation}
G(j/q+x,r)=\frac{\pi^2}{q^2 \sin^2(\pi r/q)} e^{2\pi i r j/q}+\frac{2\pi^2 i x}{q\sin(\pi r/q)}e^{i\pi r (2j+1) /q}\, .
\label{function_G} 
\end{equation} 
In the same way one proves  that function $F(x,r)$ is a piece-wise quadratic function
\begin{equation}
F(j/q+x,r)=\frac{\pi^3\cos(\pi r/q)}{q^3 \sin^3(\pi r/q)} e^{2\pi i r j/q}+\frac{2\pi^3 i x}{q^2\sin^2(\pi r/q)}e^{2\pi i r j/q}-
\frac{2\pi^3 x^2}{q \sin(\pi r/q)} e^{i\pi r (2j+1) /q}\, .
\label{function_F}
\end{equation}
In all these formulas  $j=0,\ldots,q-1$ and $0\leq x\leq 1/q$. 

\begin{figure}
\begin{minipage}{.49\linewidth}
\begin{center}
\includegraphics[width=.99\linewidth]{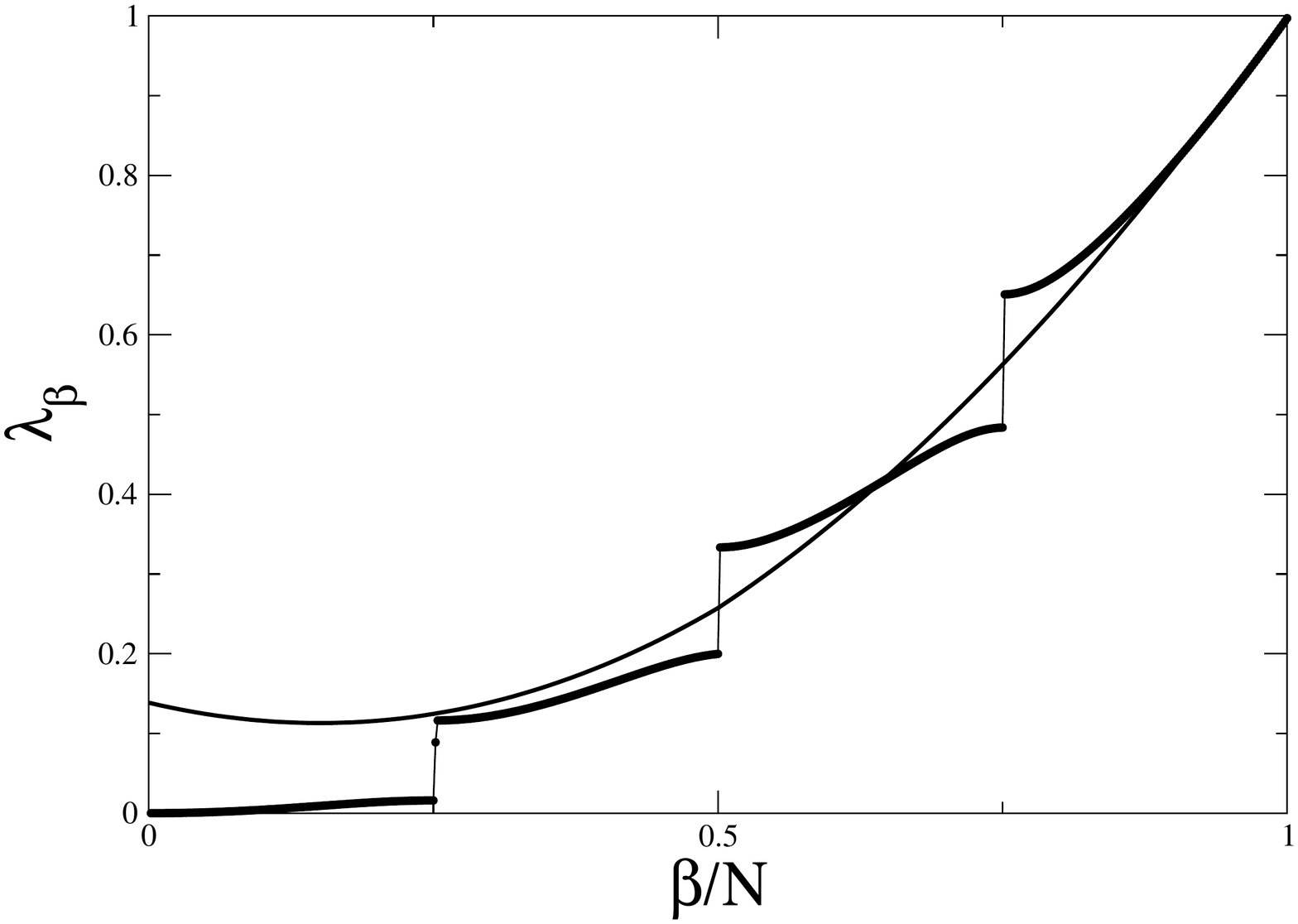}\\
(a)
\end{center}
\end{minipage}
\begin{minipage}{.49\linewidth}
\begin{center}
\includegraphics[width=.99\linewidth]{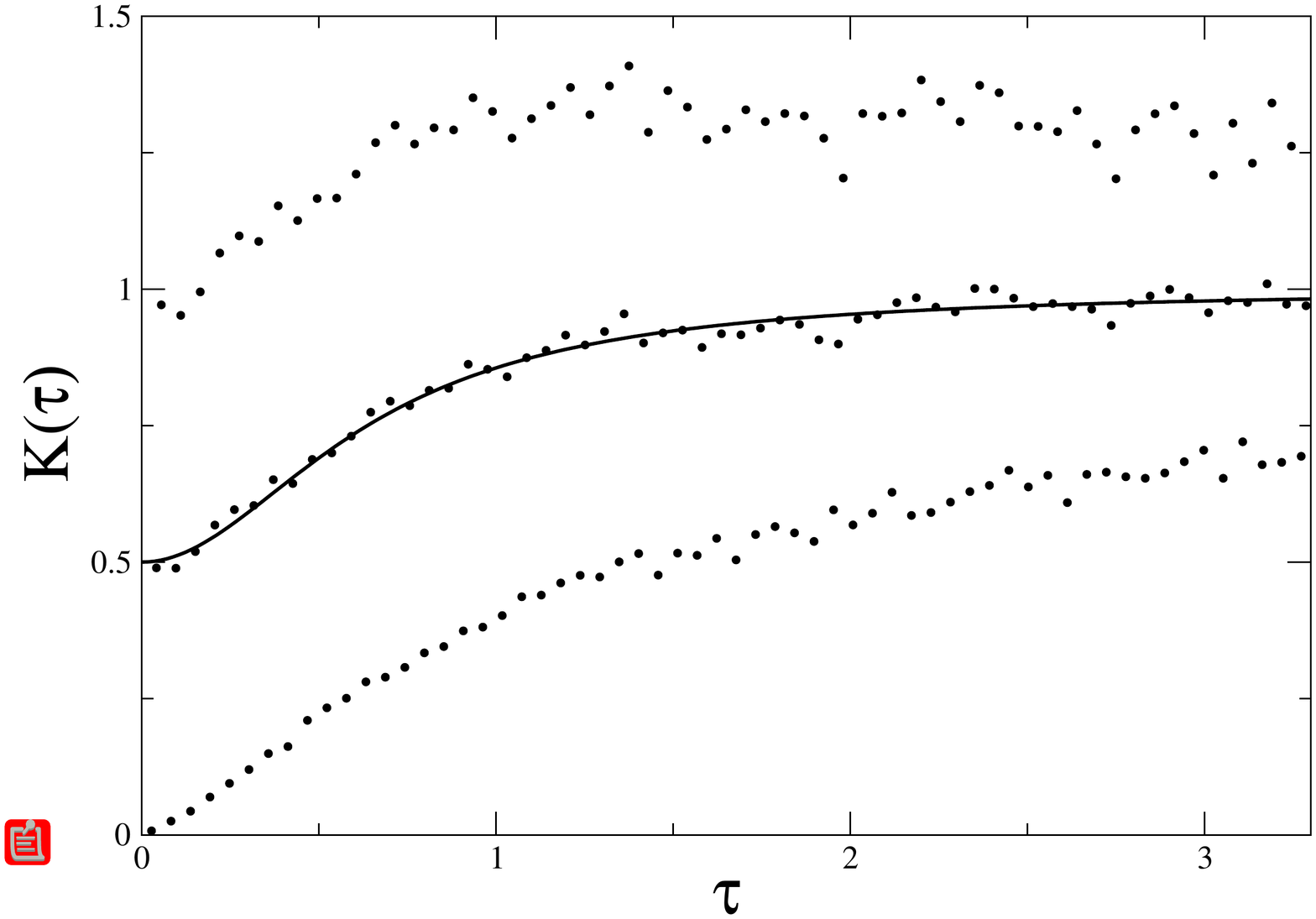}\\
(b)
\end{center}
\end{minipage}
\caption{(a)  Eigenvalues of the $P$-matrix \eqref{P_diagonal_resonant}, \eqref{P_off_diagonal_resonant} for  $h/b=2/5$ with $N_1=520$.   Solid line is the spectrum \eqref{approximate_lambda} of the asymptotic matrix.    (b) Form factor for $h/b=2/5$, $b=\pi$, $k=250.5$,and $N=400$ averaged over 1000 realisations.  Other notations are as in figure~\ref{transition_symmetric}(b). } 
\label{eigenvalues_2_5}
\end{figure}

Combining all terms together one finds  
\begin{eqnarray}
& &\lambda_{\beta}=\alpha_1 +\frac{2\pi^2 \alpha_2}{q^2} B_2\big (\big \{x(q-1)\big \}\big )+2 \mathrm{Re}\, \sum_{t=1}^{q-1} \Big (1-\frac{t}{q-1}\Big )\times \nonumber\\
&\times&e^{ixt/q}  \left [\big (\alpha_3 +\alpha_4 \cos(2\pi p t/q)\big )  G\Big (\frac{x(q-1)}{q}  ,t\Big)  +\alpha_5 \sin(2\pi p t/q)  F\Big (\frac{x(q-1)}{ q} ,t\Big) \right ]\, . 
\label{approximate_lambda}
\end{eqnarray}
The main interest for the calculation of the form factor is the behaviour of the largest eigenvalues for $x$ close to zero. Using \eqref{function_G} and \eqref{function_F}  one concludes that 
\begin{equation*}
\lambda_{\beta}= C_0+C_1 \frac{\beta}{N_1^{\mathrm{(r)}}} , \qquad 0\leq \beta\ll N_1^{\mathrm{(r)}}\, .
\end{equation*}
Here 
\begin{equation*}
C_0=\alpha_1+\frac{\pi^2 \alpha_2}{3q^2}+\frac{2\pi^2}{q^2} \sum_{t=1}^{q-1}\Big (1-\frac{t}{q-1}\Big )\Big (
\frac{\alpha_3 +\alpha_4 \cos(2\pi p t/q) }{\sin^2(\pi t/q)}+\frac{\pi \alpha_5 \sin(2\pi p r/q)\cos(\pi t/q)}{q \sin^3(\pi t/q)}\Big )   
\end{equation*}
and 
\begin{equation*}
C_1 = -\frac{2\pi^2 \alpha_2 (q-1)}{2q^2}-\frac{2\pi^2 (q-1)}{q^2} \sum_{t=1}^{q-1} \Big (1-\frac{t}{q-1}\Big )
\big (\alpha_3 +\alpha_4 \cos(2\pi p t/q)\big )\, .
\end{equation*}
The sum over  residues is of  the form 
\begin{equation*}
\sum_{t=1}^{q-1} \Big (1-\frac{t}{q-1}\Big )h(t)
\end{equation*}
and (as it is easy to check) in the considered case  $h(q-t)=h(t)$. Therefore 
\begin{equation*}
\sum_{t=1}^{q-1}t\,h(t)=\sum_{t=1}^{q-1}(q-t)\,h(q-t)=\sum_{t=1}^{q-1}(q-t)\,h(t)\longrightarrow \sum_{t=1}^{q-1}t\,h(t)=\frac{q}{2}\sum_{t=1}^{q-1} h(t) \, .
\end{equation*} 
Consequently
\begin{equation*}
C_0=\alpha_1+\frac{\pi^2 \alpha_2}{3q^2}+\frac{2\pi^2 (q-2)}{q^2(q-1)}\sum_{t=1}^{q-1}\Big (\frac{\alpha_3 +\alpha_4 \cos(2\pi p t/q)}{\sin^2(\pi t/q)}+\frac{\pi \alpha_5 \sin(2\pi p r/q)\cos(\pi t/q)}{q \sin^3(\pi t/q)}\Big )   
\end{equation*}
and 
\begin{equation*}
C_1 = -\frac{2\pi^2 \alpha_2 (q-1)}{2q^2}-\frac{2\pi^2 (q-2)}{q^2} \sum_{t=1}^{q-1} 
\big (\alpha_3 +\alpha_4 \cos(2\pi p t/q)\big )\, .
\end{equation*}
Using sums indicated in Appendix~\ref{divers} and collecting all terms in the end  one finds  that
\begin{equation*}
C_0=1,\qquad C_1=-2 .
\end{equation*}
This result signifies that largest moduli eigenvalues of the transition matrix for the barrier billiard with rational ratio $h/b=p/q$ are (i) independent on values of integers $p$ and $q$ and (ii) have the same asymptotic expression as in \eqref{approx_Lambda} (taking into account that  $N^{(\mathrm{r})}\approx 2N_1^{\mathrm{(r)}}$)
\begin{equation*}
\Lambda= \pm\left ( 1-\frac{\beta}{N_1^{\mathrm{(r)}}}+\mathcal{O}\left( \frac{1}{N_1^{\mathrm{(r)}}} \right )\right )\approx  \pm\left ( 1-\frac{2\beta}{N^{\mathrm{(r)}}}\right ) \, .
\end{equation*}
As it has been explained above it implies that  (i) the form factor for barrier billiards  is different for odd and even $n$ and (ii) the spectral compressibility is exactly equal $1/2$ for all positions of the barrier 
\begin{equation*}
\chi\equiv K(0)=\frac{1}{2}\, .
\end{equation*} 
The numerical calculations  exemplified in figure~\ref{eigenvalues_2_5} confirm well these results.  

%=======================
\section{Summary}\label{conclusion} 

It is demonstrated that the method of calculation of the level compressibility proposed by G. Tanner in \cite{tanner} for chaotic systems can  successfully be applied for intermediate statistics models. The criterium discussed in \cite{tanner} states that if the difference between the dominant eigenvalue of the transition matrix $\Lambda=1$ and the second in magnitude eigenvalue is big enough then only the dominant eigenvalues contributes to the form factor and one gets the usual value of the form factor corresponding to standard random matrix ensembles. Notably, the level compressibility is zero. 

For models considered in the paper no individual transition matrix eigenvalues dominate and one has to sum over many of  them  with moduli close to $1$.  Two types of random unitary matrices  were investigated. The first corresponding  to a  quantisation of an interval-exchange map \cite{marklof}  has been discussed  in detail in \cite{map}-\cite{integrable_ensembles}.   In particular, the  values of the level compressibility  were derived. The application of the transition matrix approach for this case serves first of all to check the validity of Tanner's  method  for intermediate statistics models. It appears that interval-exchange matrices  lead to circulant transition matrices whose eigenvalues are explicitly known and all necessary sums are easily estimated. In the end one gets the same values of the level compressibility as obtained in \cite{map}-\cite{integrable_ensembles}  but with much simpler and transparent calculations.  An example is of a special interest. It corresponds to  interval-exchange matrices with an irrational value of a parameter (which strictly speaking  describes not an interval-exchange map but only a parabolic one). Numerically it has observed in \cite{general_case} that in such case spectral statistics is of usual random matrix type (GOE or GUE depending on a symmetry) as for chaotic systems which looks strange as the Lyapunov exponent of any parabolic  map is zero.   The transition matrix approach clearly indicates that, though there is no dominant eigenvalue as been discussed in \cite{tanner},  all eigenvalues except $\Lambda=1$ for large matrix dimensions are so quickly oscillating  that averaging over a small interval of the argument  effectively removes  their contributions producing the standard random matrix result.   

The main part of the paper contains  the calculation of the level compressibility for random unitary matrices derived from the exact quantisation of barrier billiards in \cite{billiard_1,billiard_2}.  The importance of such matrices comes from the fact that  they have the same spectral statistics as high-excited states of barrier billiards which are the simplest examples of pseudo-integrable models for which very little is known analytically.  

The barrier billiard transition matrices  are more complicated that the ones for interval-exchange matrices. Their spectra contain forbidden zones and their exact eigenvalues, seems, not to be accessible in closed form.  Nevertheless, as the level compressibility requires the control only  of largest moduli eigenvalues of the transition matrix  it is possible to find such eigenvalues  for large matrix dimensions precisely. The main simplification comes from the fact that eigenvectors corresponding to largest moduli eigenvalues  are slow oscillating functions. Therefore quickly oscillating terms in matrix elements will give negligible contributions on these eigenvectors and one can substitute instead of exact matrix elements  their average over fast oscillations.  The resulting matrices are simpler and permit to find their large moduli eigenvalues analytically. In the end one proves that the level compressibility of barrier billiards for all positions and heights of the barrier is  the same and equals $1/2$. This result strongly indicates that spectral statistics of the $B$-matrices associated with barrier billiards is universal (i.e., independent on the barrier position) and well described by the semi-Poisson distribution.

%=======================

\appendix

%----------------------------------------

\section{Approximate expression for the $B$-matrix for the symmetric billiard}\label{direct_calculation} 

The purpose of this Appendix is the determination of  the transition matrix for the symmetric case (i.e., $x_m= (-1)^{m+1}y_m$, $y_m=b\sqrt{k^2-\pi^2 m^2/b^2}$)  in the paraxial approximation by taking into account only the pole terms in the definition \eqref{B_matrix}. From \eqref{L_m} with  odd $m=2j-1$ it follows (for simplicity it is assumed   $N$ is even $N=2n$ and the products is taken from $q=1$ till $q=n$)
\begin{eqnarray*}
& &2y_{2j-1} L_{2j-1}^2=(2y_{2j-1})^2  \prod_{q\neq j} \frac{y_{2j-1}+y_{2q-1}}{y_{2j-1}-y_{2q-1}} \prod_{q} \frac{y_{2j-1}-y_{2q}}{y_{2j-1}+y_{2q}}\nonumber\\
& =&\prod_{q} \Big ( \frac{y_{2j-1}+y_{2q-1}}{y_{2j-1}+y_{2q}}\Big )^2  \frac{\pi^2 \prod_{q}[q^2-(j-1/2)^2]}{ \prod_{q\neq j} [(q-1/2)^2-(j-1/2)^2]}= g_{j}\, F^2 (y_{2j-1})
\end{eqnarray*}
where
\begin{equation*}
 g_j=\lim_{v\to j-1/2} \frac{\pi^2((j-1/2)^2-v^2) \prod_{q}[1-(j-1/2)^2/q^2]}{ \prod_{q} [1-v^2/(q-1/2)^2]}, \qquad F(z)=\prod_{q} \frac{q(z+y_{2q-1})}{(q-1/2)(z+y_{2q})}\, .
\end{equation*}
Exactly in the same way one gets
\begin{eqnarray*}
& &2y_{2j} L_{2j}^2=-(2y_{2j})^2  \prod_{q\neq j} \frac{y_{2j}+y_{2q}}{y_{2j}-y_{2q}} \prod_{q} \frac{y_{2j}-y_{2q-1}}{y_{2j}+y_{2q-1}}\nonumber\\
& =&-\prod_{q} \Big ( \frac{y_{2j}+y_{2q}}{y_{2j}+y_{2q-1}}\Big )^2  \frac{\, \pi^2 \prod_{q}[(q-1/2)^2-j^2]}{ \prod_{q\neq j} [q^2-j^2]}=\frac{f_j}{F^2(y_{2j})}
\end{eqnarray*}
with
\begin{equation*}
f_j=-\lim_{v\to j} \frac{\pi^2 (j^2-v^2)\prod_{q}[1-j^2/(q-1/2)^2]}{ \prod_{q} [1-v^2/q^2]}\, .
\end{equation*}
All products in the above expressions should be taken from $1$ to $n$. If $j$ is not too close to $kb/\pi$ (i.e., the momentum is not close to the threshold of new propagating modes) the products in $g_j$ and $f_j$ can be extended to infinity and these functions can easily be calculated  from standard expressions
\begin{equation*}
\sin \pi x=\pi x \prod_{q=1}^{\infty}\Big (1-\frac{x^2}{q^2} \Big ), \qquad \cos \pi x=\prod_{q=1}^{\infty} \Big (1-\frac{x^2}{(q-1/2)^2} \Big )\, .
\end{equation*} 
In this way one finds that  $g_j=2$ and  $ f_j=2\pi^2 j^2$. Finally
\begin{equation*}
L_{2j-1}=\frac{F(y_{2j-1})}{\sqrt{y_{2j-1}}} ,\qquad L_{2j}=\frac{\pi j}{F(y_{2j}) \sqrt{y_{2j}}}\, . 
\end{equation*}
The $T$-matrix elements are 
\begin{equation*}
T_{m,n}=\frac{L_m^2 L_n^2}{(x_m+x_n)^2}\, .
\end{equation*}
In the paraxial approximation one should take into account only the terms with $x_m$ of different signs. For symmetric billiard it means that $T_{2n,2m}\approx 0$, $T_{2n-1,2m-1}=0$ and 
\begin{equation*}
T_{2m-1,2n}=\frac{L_{2m-1}^2 L_{2n}^2}{(y_{2m-1}-y_{2n})^2}\approx \frac{F^2(y_{2m-1})}{F^2(y_{2n})}\frac{\pi^2 n^2}{y_{2m-1}y_{2n}(y_{2m-1}-y_{2n})^2}\, .
\end{equation*}
As only the pole terms are important one can put $y_{2n}\approx y_{2m-1}$ and  $F^2(y_{2m-1})\approx F^2(y_{2n})$ 
\begin{eqnarray}
T_{2m-1,2n}&\approx &\frac{ \pi^2 n^2}{y_{2m-1}y_{2n}(y_{2m-1}-y_{2n})^2}=\frac{ n^2}{\pi^2 (n^2-(m-1/2)^2)^2} \frac{(y_{2m-1}+y_{2n})^2 }{ y_{2m-1}y_{2n}} \nonumber \\
&\approx& \frac{ 4 n^2}{\pi^2 (n^2-(m-1/2)^2)^2} =\frac{1}{\pi^2(n-m+1/2)^2}+ \frac{1}{\pi^2(n+m-1/2)^2}\, .
\label{T_limit}
\end{eqnarray}

%-------------------------------------

\section{Instantaneous approximation for the resonance case}\label{resonance}

When the ratio $h/b$ is a rational number  $h/b=p/q$ with co-prime integer $p$ and $q$ it is plain that for the barrier billiard as in figure~\ref{barrier_fig}(a) the following 3 transverse momenta with integer $t$ (and the corresponding longitudinal momenta $p_t=\sqrt{k^2-\pi^2 w_t^2}$) are equal
\begin{equation}
w_t=\frac{\pi qt}{b}=\frac{\pi ((q-p)t}{b-h}=\frac{\pi p t}{h}\, .
\label{resonance_momenta} 
\end{equation}
Introduce the elementary solutions with these momenta  in each of 3 regions indicated in figure~\ref{barrier_fig}(b) 
\begin{eqnarray*}
\phi_{qt}^{(1)}(x,y)&=&\sqrt{\frac{2}{b}} \sin\Big (\frac{\pi q t }{b} y\Big )\exp \big( i p_t  x\big )\, ,
\qquad  0\leq y\leq b ,\qquad x<0 \, ,\nonumber \\
\phi_{(q-p)t}^{(2)}(x,y)&=&\sqrt{\frac{2}{b-h}} \sin\Big (\frac{\pi (q-p) t }{b-h}(b- y)\Big )\exp \big(i p_t x\big ) \, ,
\qquad  b-h\leq y\leq b,\qquad x>0\, ,\nonumber \\
\phi_{pt}^{(3)}(x,y)&=&\sqrt{\frac{2}{h}} \sin\Big (\frac{\pi p t }{h} y\Big )\exp\big ( i p_t x \big )\, ,
\qquad  0\leq y\leq h,\qquad x>0 \, .
\end{eqnarray*}
Due to the resonant conditions \eqref{resonance_momenta} all these solutions represent  exact solutions for the scattering inside the slab in figure~\ref{barrier_fig}(b).  The number of such solutions is
\begin{equation}
N_0=\left [\frac{k b}{\pi q}\right ].
\label{N_0}
\end{equation}
 When spectral statistics of  non-trivial eigenvalues is considered these solutions should be removed.  It has been done in detail in \cite{billiard_2}. Below the derivation of the paraxial approximation for the $T$-matrix in such case is briefly  discussed. 

The paraxial approximation for  $T$-matrix for the scattering inside the slab in figure~\ref{barrier_fig}(b) for non-resonant waves when in the first region $n_1\neq 0\  \mathrm{mod}\ q$, in the second region $n_2\neq 0\mod q-p$,  and in the third one $n_3\neq 0\mod p$ are given by the same expression as in \eqref{t_paraxial}. The first step consists in removing all wave from region $1$ proportional to $q$. But it is not enough as waves from the second and the third regions can diffract into waves in the first region with $n_1\equiv 0 \mod q$. To remove them notice that 
\begin{equation*}
\phi_{qt}^{(1)}(x,y)=\left \{ \begin{array}{cc} -(-1)^{qt}  \sqrt{\dfrac{b}{b-h}} \phi_{(q-p)t}^{(2)}(x,y),&  h\leq y\leq b\\
 \sqrt{\dfrac{b}{h}} \phi_{p t}^{(3)}(x,y),& 0\leq y\leq b \end{array} \right . . 
\end{equation*}
Therefore the following linear combination \cite{billiard_2} 
\begin{equation*}
\phi_t^{(4)}(x,y)= \sqrt{\frac{h}{b}}  \phi_{(q-p)t}^{(2)}(x,y) +(-1)^{qt} \sqrt{\frac{b-h}{b}}  \phi_{(q-p)t}^{(3)}(x,y)
\end{equation*}
is orthogonal to $\phi_{qt}^{(1)}(x,y)$ and cancels undesirable waves with $n_1\equiv 0 \mod q$.  

The calculation of the scattering into such state can be performed as above
\begin{eqnarray*}
& & S_{m,t}^{1\to 4}=2\sqrt{\frac{h}{b^2(b-h)}}\int_h^b \sin\Big (\frac{\pi m y}{b} \Big ) \sin\Big (\frac{\pi (q-p)t(b- y)}{b-h}\Big )dy \nonumber\\
&+&
2(-1)^{q t}\sqrt{\frac{b-h}{b^2 h} } \int_0^h  \sin\Big (\frac{\pi m y}{b}\Big ) \sin\Big (\frac{\pi p t y}{h}\Big ) dy \nonumber\\
&=& \frac{(-1)^{(p+q)t}\sin\big (\pi m h/b\big )  b}{\sqrt{h(b-h)}} \Big ( \frac{1}{m-q t } -\frac{1}{m+qt } \Big )\, .
\end{eqnarray*}
The paraxial approximation of the corresponding transition matrix elements  is given by the pole term 
\begin{equation}
t_{m,t}^{1\to 4} =\frac{\sin^2(\pi m h/b)}{\pi^2 p (q-p) (m/q-t)^{2}}, \qquad m\not\equiv 0\ \mod q\, . 
\label{t_2_3}
\end{equation}
%-------------------------------------
\section{Divers relations}\label{divers}

In this Appendix a few  formulas used in the text are briefly reviewed.

The known identity (see e.g., \cite{bateman}) 
\begin{equation}
 \sideset{}{'}\sum_{r=-\infty}^{\infty} \frac{1}{r^n}e^{2\pi irx}=-\frac{(2\pi i)^n}{n!}B_n(\{x\})
  \label{Bernoulli}
\end{equation} 
where $\{x\}$ is the fractional part of $x$ and $B_n(x)$ are Bernoulli polynomials. For example,  
\begin{equation}
B_2(x)=x^2-x+\frac{1}{6}\, ,\qquad B_3(x)=x^3-\frac{3}{2}x^2+\frac{1}{2}x\, .
\label{B_2_3}
\end{equation}
The following identities are standard and  presented for completeness.
 \begin{equation}
\sum_{n=-\infty}^{\infty}\frac{1}{n+x}=\pi \cot(\pi x),\qquad \sum_{n=-\infty}^{\infty}\frac{1}{(n+x)(n+y)}=\frac{\pi\big (\cot(\pi x)-\cot(\pi y)\big )}{y-x}\, .
\label{standard_formulas}
\end{equation}
The first formula is simply the expansion of the right-hand side over poles. The second is a consequence of the first. Differentiating  the above expressions  by $x$ and $y$ shows  that
\begin{eqnarray}
\sum_{n=-\infty}^{\infty}\frac{1}{(n+x)^2}&=&\frac{\pi^2}{\sin^2(\pi x)},\qquad 
\sum_{n=-\infty}^{\infty}\frac{1}{(n+x)^4}=\frac{\pi^4 (3-2\sin^2(\pi x))}{3\sin^4(\pi x)},\label{useful_formulas}\\
\sum_{n=-\infty}^{\infty}\frac{1}{(n+x)^2(n+y)^2}&=&\frac{\pi^2}{(x-y)^2}\Big (\frac{1}{\sin^2(\pi x)} +\frac{1}{\sin^2(\pi y)} \Big ) +\frac{2\pi (\cot(\pi x)-\cot(\pi y))}{(x-y)^3}\,  . \nonumber
\end{eqnarray}
In different places of the paper one needs to calculate finite sums  $\sum_{t=1}^{q-1} f(t)$  over residues $\mod q$.
A characteristic feature of such sums is that their summands $f(t)$  can be rewritten as ratio of two polynomials in variable $z=e^{2\pi i t/q}$. The summation over $t$ from $1$ to $q-1$ corresponds to the calculation of the integral 
\begin{equation*}
\sum_{t=1}^{q-1} f(t)=\frac{1}{2\pi i} \oint_C \frac{q z^{q-1}}{z^q-1} f(z)dz
\end{equation*}
where contour $C$ encircled all roots of $z^q-1=0$ except the one with $z=1$. By deforming the contour and calculating the necessary residues one can obtain the necessary sums in closed form.  
Below  a  few  formulas obtained by this manner are listed
\begin{eqnarray}
& &\sum_{t=1}^{q-1}\frac{1}{\sin^2(\pi t/q)}=\frac{q^2-1}{3}\, ,\qquad 
 \sum_{t=1}^{q-1}\frac{\cos(2\pi p t/q)}{\sin^2(\pi t/q)}=\frac{q^2+6p^2-1}{3}-2 q p \, ,\label{F_1}\\
& & \sum_{t=1}^{q-1}\frac{\sin(2\pi p t/q)\cos(\pi t/q)}{\sin^3(\pi t/q)}=\frac{2p(q^2+2p^2)}{3}-2 q p^2\, ,\\
& & \sum_{m=1}^{q-1}\frac{\sin^2(\pi p m /q)\sin^2(\pi p (m+r)/q)}{\sin^2(\pi m/q)}=\frac{pq}{2}-p^2 \sin^2(\pi p r/q)\, ,\\ 
& & \sum_{m=1}^{q-1}\frac{\sin^4(\pi p m /q)}{\sin^4(\pi m/q)}=\frac{pq(2p^2+1)}{3}-p^4\, .\label{F_4}  
\end{eqnarray}
Here it is assumed that $p/q<1/2$. 

In Section~\ref{rational_BB} one has to calculate the following sum where $\nu(k)$ is the $k^{\mathrm{th}}$ integer not divisible by $q$
\begin{equation*}
S=\frac{1}{N}\sum_{j,k=1}^N  f(\nu(j)-\nu(k))e^{ix(j-k)}
\end{equation*}
in the limit $N\to\infty$ with a certain quickly decreasing function $f(x)$, $f(x)\underset{|x|\to \infty}{\longrightarrow} x^{-2}$. 

To get an explicit expression of that sum  notice that  the number of integers from $1$ to $n$ divisible by $q$ is $[n/q]$ where $[x]$ is the largest integer less of equal $n$. Therefore if $n=q l+r$ with $r=1,\ldots,q-1$ then $k=(q-1)l+r$. It means that
\begin{equation}
\nu((q-1)l+r)=q l +r\, ,\qquad r=1,\ldots, q-1\, . 
\label{nu}
\end{equation}
As integer $\nu(k) \neq 0 \mod q$ the residue $r\neq 0$. 
 
Writing $j=(q-1)m+s$ and $k=(q-1)n+r$ with integer $m$, $n$  and $r,s=1,\ldots, q-1$ one gets that 
\begin{equation*}
j-k=(q-1) u+t,\qquad \nu(j)-\nu(k)=q u + t, \qquad  u=m-n, t=s-r\, .
\end{equation*}
The summation over integers $j$ and $k$ is equivalent to the summation over  integers $m,n$ and $s,r$. Fixing the differences $k=m-n$ and $t=s-r$,  using the fact that integers with fixed residue $\mod (q-1)$ are uniformly distributed 
\begin{equation*}
\lim_{N\to\infty} \frac{1}{N}\sum_{j=1}^N F(j\mod (q-1))=\frac{1}{q-1}\sum_{s=1}^{q-1}F(s) 
\end{equation*}
and that 
\begin{equation*}
\sum_{s=1}^{q-1}\sum_{r=1}^{q-1}f(s-r)=\sum_{t=-(q-2)}^{q-2}(q-1-|t|) f(t)
\end{equation*}
one finds that 
\begin{equation*}
S=\sum_{u=-\infty}^{\infty}\sum_{t=-(q-2)}^{q-2}\left (1-\frac{|t|}{q-1} \right ) f(q u+t)e^{ix((q-1)u+t)}\, .
\end{equation*}
Due to a quick decrease of function $f(x)$ the summation over $u$ can be  extended to the sum over all integers. It is convenient to separate term with $t=0$, add together terms with $t$ and $-t$, and in the last term change $u\to-u$
\begin{eqnarray*}
S&=&\sum_{u=-\infty}^{\infty}f(qu)e^{ix(q-1)u} \nonumber \\
&+&\sum_{u=-\infty}^{\infty}\sum_{t=1}^{q-1}\left (1-\frac{t}{q-1} \right )\left (  f(q u+t)e^{ix((q-1)u+t)}+f(-q u-t)e^{-ix((q-1)u+t)}\right )\, .
\end{eqnarray*}
The used  function $f$ is even $f(-x)=f(x)$ and this expression can be written as follows
\begin{equation}
S=\sum_{u=-\infty}^{\infty}f(qu)e^{ix(q-1)u} 
+2\mathrm{Re}\,\sum_{u=-\infty}^{\infty}\sum_{t=1}^{q-1}\left (1-\frac{t}{q-1} \right )  f(q u+t)e^{ix((q-1)u+t)}\, .
\label{sum_S}
\end{equation}

%=======================

\end{document}